\documentclass[preprintnumbers,11pt,a4paper,oneside]{article}
\pdfoutput=1

\usepackage{jheppub}
\usepackage{color}
\usepackage{graphicx}
\usepackage{wrapfig,enumerate,slashed}

\usepackage{footmisc}
\usepackage{amsmath}
\usepackage{wasysym} 
\usepackage{graphicx}
\usepackage{subcaption}
\usepackage{color}
\usepackage{comment}
\usepackage{appendix}
\usepackage{slashed}

\newcommand{\be}{\begin{eqnarray}}
\newcommand{\ee}{\end{eqnarray}}

\newcommand{\bee}{\begin{eqnarray}}
\newcommand{\eee}{\end{eqnarray}}
\newcommand{\beeq}{\begin{equation}}
\newcommand{\eeeq}{\end{equation}}

\usepackage{tikz}
\usepackage{etoolbox}
\newcommand{\circled}[2][]{%
  \tikz[baseline=(char.base)]{%
    \node[shape = circle, draw, inner sep = 1pt]
    (char) {\phantom{\ifblank{\smaller \smaller #1}{\smaller \smaller #2}{\smaller \smaller #1}}};%
    \node at (char.center) {\makebox[0pt][c]{\smaller \smaller #2}};}}
\robustify{\circled}

\usepackage[ruled,vlined]{algorithm2e}
\usepackage{setspace}
\AtBeginEnvironment{algorithm}{\setstretch{1.2}}
\makeatletter
\renewcommand*{\@algocf@post@ruled}{}
\makeatother
\makeatletter
\renewcommand*{\@algocf@pre@ruled}{}
\makeatother
\makeatletter
\renewcommand*{\@algocf@capt@ruled}{}
\makeatother
 
\makeatletter
\gdef\@fpheader{}
\makeatother
\begin{document}

\title{Collider Events on a Quantum Computer}

\author[a]{Gösta Gustafson,}
\author[a]{Stefan Prestel,}
\author[b]{Michael Spannowsky,}
\author[c]{Simon Williams}

\affiliation[a]{Department of Astronomy and Theoretical Physics, Lund University, S-223 62 Lund, Sweden}
\affiliation[b]{Institute for Particle Physics Phenomenology, Department of Physics, Durham University, Durham DH1 3LE, U.K.}
\affiliation[c]{High Energy Physics Group, Blackett Laboratory, Imperial College, Prince Consort Road, London, SW7 2AZ, United Kingdom}

\abstract{High-quality simulated data is crucial for particle physics discoveries. Therefore, parton shower algorithms are a major building block of the data synthesis in event 
generator programs. However, the core algorithms used to generate parton showers 
have barely changed since the 1980s.
With quantum computers' rapid and continuous development, dedicated algorithms
are required to exploit the potential that quantum computers provide to address problems
in high-energy physics. This paper presents a novel approach to synthesising parton showers using the Discrete QCD method. The algorithm benefits from an elegant quantum walk implementation which can be embedded into the classical toolchain. 
We use the \texttt{ibm\_algiers} device to sample parton shower configurations and 
generate data that we compare against measurements taken at the ALEPH, DELPHI 
and OPAL experiments. This is the first time a Noisy Intermediate-Scale Quantum (NISQ) device has been used to 
simulate realistic high-energy particle collision events.}

\preprintA{IPPP/22/43}
\preprintB{LU-TP-22-49}

\maketitle

\section{Introduction}

Particle physics studies the smallest building blocks of matter and their interactions by analysing the outcome of high-energy particle scattering events produced in large-scale particle colliders. In such collision events, the colliding particles annihilate, and the resulting energy often creates highly energetic particles. These short-lived states quickly decay into high occupation-number (i.e. large particle multiplicity), low average-energy states through a process called ``showering". This phenomenon is particularly important for colour-charged partons, such that developing ``parton shower" calculations~\cite{Fox:1979ag,Sjostrand:1985xi,Gustafson:1986db} have become a crucial aspect of research in Quantum Chromodynamics (QCD)~\cite{Lonnblad:1992tz,Gieseke:2003rz,Sjostrand:2004ef,Schumann:2007mg,Platzer:2009jq,Nagy:2014mqa,Hoche:2015sya,Fischer:2016vfv,Hamilton:2020rcu}.  

All processes in particle physics are inherently quantum mechanical, making particle physics an excellent driver for quantum algorithm development, as pointed out early on by Feynman~\cite{Feynman:1981tf}. One major obstacle for such applications is tackling environmental noise within the device, which can cause decoherence effects, drastically reducing the fidelity of quantum computers. Rapid progress in manufacturing and running Noisy Intermediate-Scale quantum (NISQ) devices has led to renewed interest in particle physics applications~\cite{Jordan:2014tma,Garcia-Alvarez:2014uda,Arrighi:2018PRA,Marque-Martin:2018PRA,Alexandru:2019nsa,Jay:2019PRA,Wei:2019rqy,Lamm:2019bik,MottQuantum,Bauer:2019qxa, Alexandru:2019ozf,Blance:2020ktp,Lamm:2019uyc,Abel:2020ebj,Abel:2020qzm,Bepari:2020xqi,Williams:2021lvr,DiMolfetta:2020QIP,Araz:2022haf,Matchev:2020wwx,DeJong:2020riy,Ngairangbam:2021yma,Bauer:2022hpo,Agliardi:2022ghn}, allowing to test the limits of current quantum hardware. However, these applications are, so far, limited to proof-of-principle studies, often focusing on implementations of low-dimensional field theories or simplified models of quantum circuits. 

The QCD parton showering process is crucial in modelling and analysing data from particle colliders. This success of parton shower programs is based on the ability to produce ``synthetic" scattering event data, containing physically meaningful particle final states that obey momentum conservation~\cite{Buckley:2011ms}. Such simulated data can be compared to actual scattering event data recorded by particle detectors, hinting at necessary improvements in the theory prediction -- or discoveries. Due to its relevance, the showering process has also generated interest in the quantum computing community~\cite{Bauer:2019qxa,Bepari:2020xqi,Williams:2021lvr,Deliyannis:2022uyh}. These attempts considered the showering of individual, isolated particles. The generation of physically meaningful events was deemed impossible, in particular, since it is assumed that each particle produced by the shower samples had continuous momentum-space quantum numbers. However, the usefulness of a semi-classical picture of factorising the model of scattering events into the evolution of individual partons (inspired by the early work of Feynman and Field~\cite{Field:1977fa} and by factorisation proofs for specific, very simple one-dimensional measurements~\cite{Collins:1984kg,Collins:1989gx}) has to be discarded for the multi-dimensional modelling of scattering events that make current parton showers successful.

This note proposes the first quantum algorithm capable of simulating realistic scatting events that can be compared to data recorded at collider experiments, achieved by reframing the classical parton shower algorithm as a quantum walk~\cite{PhysRevA.48.1687, QWProc, Kempe}. The reason to attempt implementing a quantum algorithm for parton showers is twofold. First, on an abstract level, quantum algorithms might eventually help overcome classical bottlenecks. These might be related to raw computing power or overly complicated algorithms: conventional parton showering programs implement very complicated code structures when embedding quantum effects due to corrections to the ``factorisation" picture above, both for high-energy configurations and low-energy ``soft" gluons. Thus, the development of a quantum shower algorithm allows to increase also the classical ``toolbox" (which effectively consists of a single algorithm, the ``veto algorithm", see e.g.~\cite{Sjostrand:2006za,Bierlich:2022pfr} for details) and might provide more natural realisations of quantum coherence effects. More pragmatically, we also aim to assess the relevance of quantum noise on a real-world example -- the description of data collected by collider experiments. 

In pursuing this goal, we are led to several improvements over previous work. The most important aspect is the realistic handling of parton decay kinematics, defined for any momentum-space configuration. This naturally includes momentum-conservation constraints. Other important realisations are that the energy-scale dependence of the strong coupling $\alpha_s$ -- which was omitted previously -- and using soft-gluon (quantum) interference as a guiding principle for quantum parton shower developments allows clarifying the treatment of momentum-space sampling sufficiently to permit the generation of event data.

\section{The Parton Shower Algorithm}

Parton shower processes evolve high-energy few-particle states to low-energy multi-particle states by successively decaying particles into lower-energy decay products. This process is dominated by collinear decay modes, in which the decay products are almost parallel, or soft modes, which feature at least one vanishing-energy decay product. Within the former limit, successive decay steps factorise into independent quasi-classical steps, whereas interference contributions only allow for partial factorisation in the latter. Interestingly, the leading contributions to the decay rate in the collinear limit are included in the ``soft" decay rate. This makes the soft limit an excellent starting point for developing parton showers~\cite{Azimov:1985zta,Gustafson:1986db}. In this limit, the decay of a high-energy state proceeds by decaying colour-anticolour dipoles into lower-energy dipoles. Interpreting the showering process as dipole decays allows to describe the most straightforward interference patterns but also addresses momentum conservation within sequential decay processes. Individual particles that fulfil relativistic on-shell relations $p^\mu p_\mu=m^2$ cannot decay into on-shell decay products that have non-vanishing angles or energies. Colour dipoles contain two on-shell partons, which may readily decay into three on-shell particles (forming two decay-product dipoles) without violating momentum conservation. Thus, the dipole decay pattern is well-defined throughout momentum space, i.e.\ the dipole picture allows for the generation of physical multi-particle data. The ability of \emph{all} state-of-the-art conventional showers to synthesise data events rests on dipole-shower-inspired strategies to generate the decay product momenta~\cite{Lonnblad:1992tz,Gieseke:2003rz,Sjostrand:2004ef,Schumann:2007mg,Platzer:2009jq,Nagy:2014mqa,Hoche:2015sya,Fischer:2016vfv,Hamilton:2020rcu}.

Colour-charged states predominantly decay through (soft) gluon emission, the inclusive decay probability of which is given by the eikonal interference pattern
\begin{align}
d\mathcal P \left( q(p_\mathrm{I}) \bar q(p_\mathrm{K}) \rightarrow q(p_i) g(p_j) \bar q(p_k)\right) 
\simeq \frac{ds_{ij}}{s_\mathrm{IK}} \frac{ds_{jk}}{s_\mathrm{IK} } C \frac{\alpha_s}{2\pi} \frac{2s_{ik}}{s_{ij}s_{jk}} \frac{d\phi}{2\pi} 
\simeq
\frac{ds_{ij}}{s_\mathrm{IK}} \frac{ds_{jk}}{s_\mathrm{IK} } C \frac{\alpha_s}{2\pi} \frac{2s_\mathrm{IK}}{s_{ij} s_{jk}}
 \, ,
\label{eq:eikonal_prob}
\end{align}
where $C$ is a colour-charge factor, the two-particle invariants $s_{ab}$ are defined by $s_{ab}=2p_a\dot p_b$, and where the momentum conservation condition $p_\mathrm{I} + p_\mathrm{K} = p_i + p_j + p_k$ implies the relation $s_\mathrm{IK}=s_{ik}+s_{ij}+s_{jk}$ between pre- and post-branching momenta. This decay process can equally be interpreted as the decay of a dipole of invariant mass $s_\mathrm{IK}$ into two dipoles of invariant masses $s_{ij}$ and $s_{jk}$, which are linked by the soft gluon. Dipole showers employ this inclusive probability to calculate the contribution of exclusive, fixed particle-multiplicity states $\Phi$ to an observable $O(\Phi)$ according to the master equation
\begin{align}
\mathcal F_n (\Phi_n, t_n,t_c; O) 
= \Delta(t_n,t_c)O(\Phi_n) 
+ \int\limits^{t_n}_{t_c} dt d\xi \frac{d\phi}{2\pi} C \frac{\alpha_s}{2\pi} \frac{2s_{ik}(t,\xi)}{s_{ij}(t,\xi)s_{jk}(t,\xi)} 
\Delta(t_n,t ) \mathcal F_n (\Phi_{n+1}, t,t_c; O)\,
\label{eq:psoperator}
\end{align}
where the no-branching probability $\Delta$ is given by
\begin{align}
\Delta(t_n,t) 
=
\exp\left(
-
\int\limits^{t_n}_{t} dt d\xi \frac{d\phi}{2\pi} C \frac{\alpha_s}{2\pi} \frac{2s_{ik}(t,\xi)}{s_{ij}(t,\xi)s_{jk}(t,\xi)} 
\right)\,
\end{align}
and where a particular choice of the functional form of $t$ and $\xi$ is called the phase space parametrisation or momentum mapping. Finally, $t$ defines the ``evolution variable". Branchings are successively ordered in this variable, with large-$t$ branchings occurring earlier in the evolution of the state. Large evolution variables indicate transitions at high energy, while low evolution variables indicate low-energy transitions. The first term in Eq.\ \ref{eq:psoperator} models the probability of the $n$-body state $\Phi_n$ not decaying above $t_c\simeq\Lambda_\mathrm{QCD}^2\simeq (300 \mathrm{GeV})^2$, while the second term describes one or more decays. This introduction highlights that the development of a parton shower relies on $a)$ the choice of inclusive decay probabilities $d\mathcal P$, $b)$ the choice of an evolution variable $t$, and $c)$ the choice of a momentum mapping $s_{ij},s_{jk}\leftrightarrow t,\xi$ which determines the relations between pre-and post-decay momenta. 

It is worth noting that all conventional state-of-the-art parton showers use slight variations of a single algorithm -- the ``veto algorithm" -- to solve Eq.\ \ref{eq:psoperator} numerically. This algorithm treats the variables $t$ and $\xi$ as continuous degrees of freedom. It is thus unsuitable for (current) quantum devices. The following section will develop other algorithmic solutions of Eq.\ \ref{eq:psoperator}, guided by keeping in mind the feasibility of NISQ devices.

\subsection{Reinterpreting classical parton shower algorithms as random walks}\label{sec:classicalAlgo}

This section extends the classical shower algorithm toolbox by performing several abstractions of the features of dipole showers. We are led to conclude that the showering process can be described by creating and sampling from a fixed set of primitive fractal structures, followed by a translation of the chosen primitive structure into scattering event momenta. The first step has an elegant implementation on intermediate-scale quantum devices.

\begin{figure}
\centering
\includegraphics[width=0.85\textwidth]{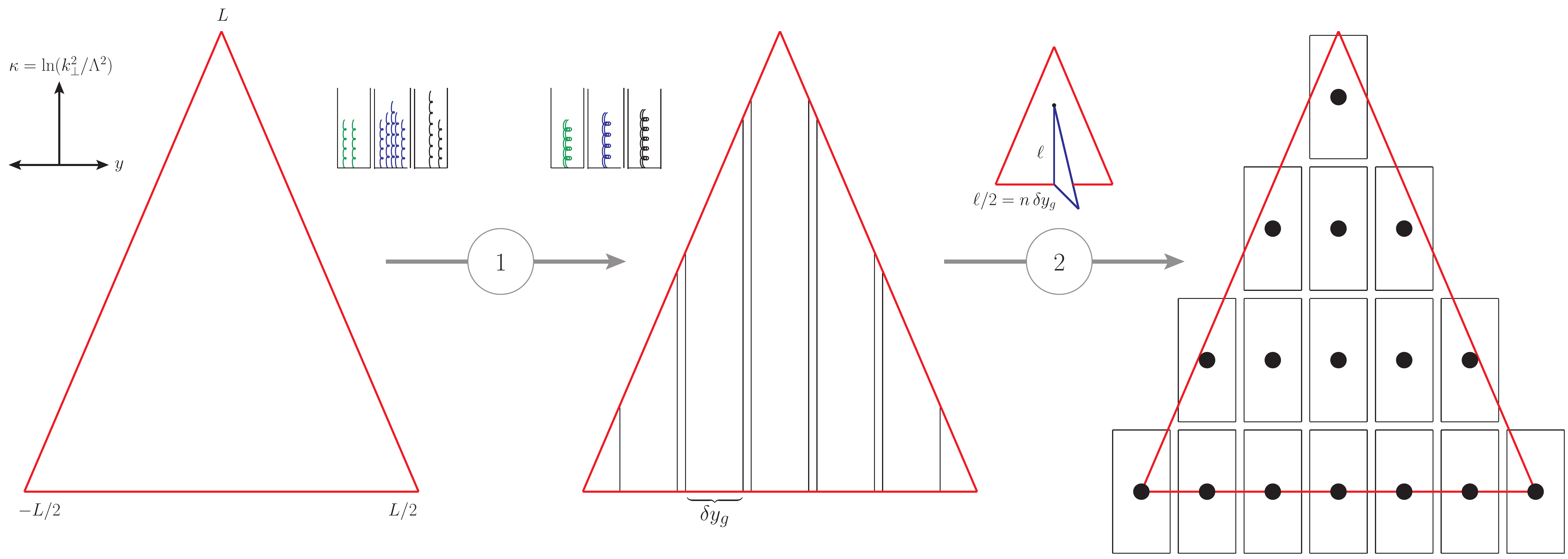}{}
\caption{\label{fig:dqcd} The phase space of effective gluon emission is discrete, since \raisebox{0.15ex}{\circled[A]{1}} gluons within a rapidity region $\delta y_g$ act coherently due to running-coupling effects. The $\kappa$ (or equivalently the $k_\perp^2$) dimension is also quantised, since \raisebox{0.15ex}{\circled[A]{2}} additional phase space folds opening due to gluon emission are quantised in units of $\delta y_g$. See main text for more details.}
\end{figure}

\begin{figure}
\centering
\includegraphics[width=0.85\textwidth]{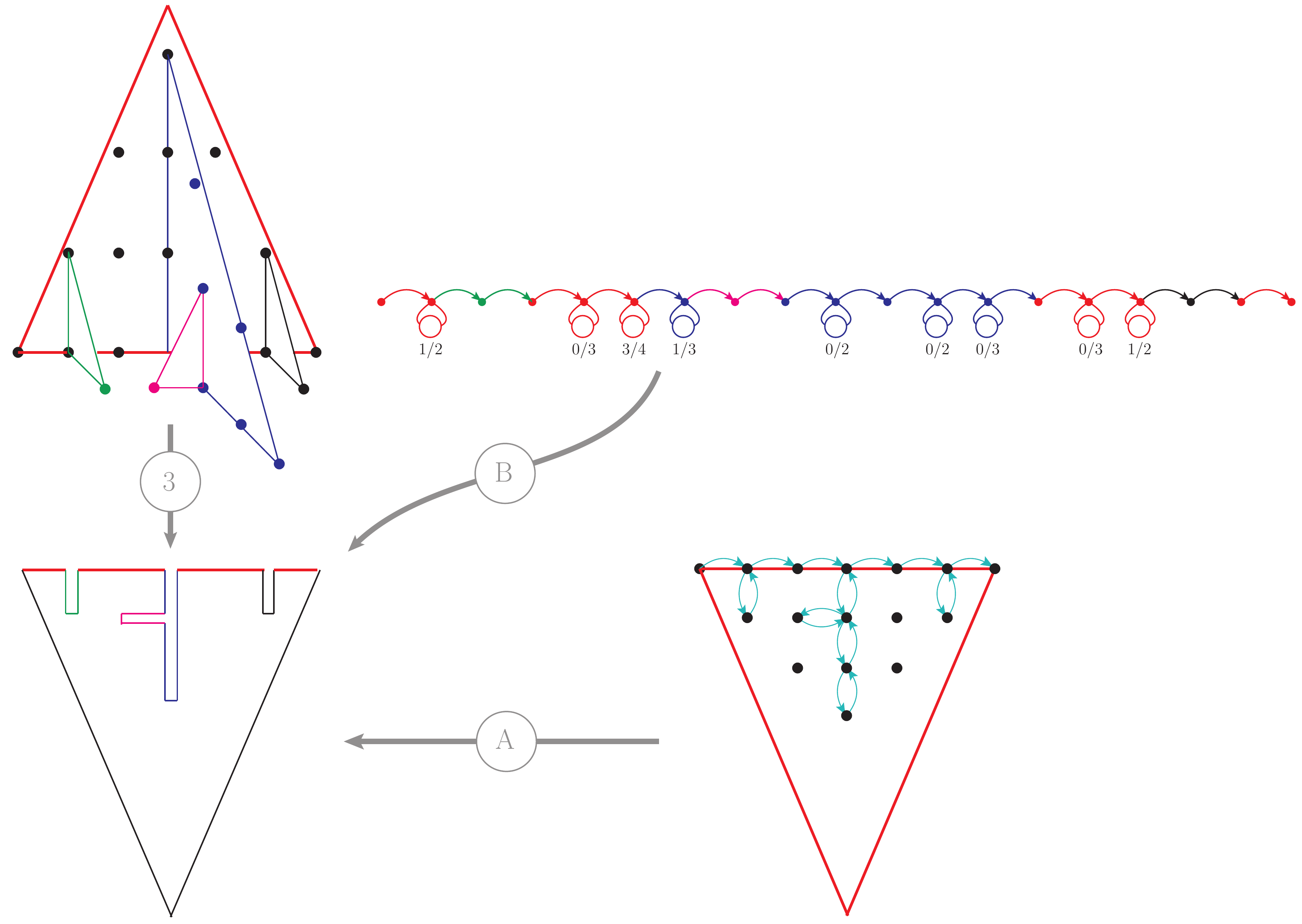}{}
\caption{\label{fig:dqcd-baseline} The Discrete QCD parton shower algorithm can be re-interpreted as a one-dimensional random walk, since \raisebox{0.15ex}{\circled[A]{3}} the baseline of the folded structure carries all necessary information. The ``grove-like" baseline structure can \raisebox{0.15ex}{\circled[A]{A}} be generated by a heavily constrained two-dimensional random walk. Due to the low fractal dimension of the grove structure, a one-dimensional random-walk algorithm \raisebox{0.15ex}{\circled[A]{B}} is equally viable. For \raisebox{0.15ex}{\circled[A]{B}}, the notation $n/n_{max}$ indicates that option $n$ was picked out of $n_{max}$ choices. The two-particle invariants ($\ln(s_{ij})$) can be read off by following the path from particle $i$ to particle $j$, and skipping segments whose colour was created and reabsorbed along the way (e.g.\ skipping the pink segments when calculating the invariant mass of the green and blue gluon tips).}
\end{figure}

The first abstraction to consider is removing the independent treatment of decay probability and momentum-space integration by absorbing the non-uniform probability density in Eq.\ \ref{eq:eikonal_prob} into the integration measure. This can be obtained by choosing a phase-space parametrisation in terms of the gluon's transverse momentum,
\begin{align}
k_\perp^2 = \frac{s_{ij} s_{jk}} {s_\mathrm{IK}} \quad \textnormal{and rapidity}\quad
y = \frac{1}{2} \ln\left( \frac{s_{ij}}{s_{jk}}\right)\, ,
\end{align}
which leads to
\begin{align}\label{eq:probspace}
d\mathcal P \left( q(p_\mathrm{I}) \bar q(p_\mathrm{K}) \rightarrow q(p_i) g(p_j) \bar q(p_k)\right) 
\simeq
=
\frac{C\alpha_s}{\pi} d \kappa dy
 \, \quad \textnormal{with} \quad \kappa = \ln \left(k_\perp^2/\Lambda^2)\right)\, ,
\end{align}
where $\Lambda^2$ is an arbitrary mass scale. Within this phase space parametrisation, allowed dipole decays are constrained to a triangular region of height $L=\ln(s_\mathrm{IK}/\Lambda^2)$ in the $(y,\kappa)$-plane, as illustrated by the left-hand panel of Fig.\ \ref{fig:dqcd}. Due to the colour charge of an emitted gluon, the rapidity span for subsequent dipole decays (at lower $\kappa$) is increased with respect to\ the originally allowed range. This feature of QCD can be interpreted as ``folding out" smaller triangular regions of additional phase space from the original triangle. Since these regions may again contribute to allowed decays, this process may repeat, leading to a fractal structure of triangles-attached-to-triangles. An example structure is shown in the upper left panel of Fig.\ \ref{fig:dqcd-baseline}. This fractal picture is the starting point of successful conventional dipole showers such as \textsc{Ariadne}~\cite{Lonnblad:1992tz,Gustafson:1987rq}.

These abstractions already simplify the treatment of parton showering since complicated ``splitting functions" have been subsumed into the phase space parametrisation. Moreover, for fixed coupling $\alpha_s$, the inclusive decay rate is uniformly distributed in the $(y,\kappa)$-plane, allowing for straightforward sampling algorithms. Nevertheless, this continuous-variable dipole decay picture is not yet suited for current universal quantum devices. 

Examining the effect of a transverse-momentum-dependent running coupling -- as supported by higher-order QCD calculations -- provides a path to an even simpler picture. We may write
\begin{align}
\alpha_s(k_\perp^2)
= \frac{12\pi}{33-2n_f} \frac{1}{\ln\,(k_\perp^2/\Lambda_\mathrm{QCD}^2)}\,,
\end{align}
where the $n_f$-dependent term arises through $g\rightarrow q\bar q$ splitting. Neglecting the latter and combining the expression with Eq.~\ref{eq:probspace} leads to
\begin{align}
d\mathcal P \left( q(p_\mathrm{I}) \bar q(p_\mathrm{K}) \rightarrow q(p_i) g(p_j) \bar q(p_k)\right) 
\simeq
=
\frac{d \kappa}{\kappa} \frac{dy}{\delta y_g}\quad\textnormal{with}\quad \delta y_g=\frac{11}{6}\,,
\label{eq:dqcd-inc-rate}
\end{align}
and where we have used that in the leading-colour limit $C\rightarrow C_\mathrm{A}/2=3/2$ for any dipole decay. As argued in~\cite{Andersson:1995jv}, interpreting the running-coupling renormalisation group equation as gain-loss equation means that gluons within a rapidity range $\delta y_g$ act coherently as one ``effective gluon", as illustrated by \raisebox{0.15ex}{\circled[A]{1}} in Fig.\ \ref{fig:dqcd}. Thus, the rapidity range of each triangular phase space region is {\it quantised} into multiples of $\delta y_g$. Since the baseline of an additional triangle extends to positive $y$ by $\ell/2$, the height $\ell$ at which to emit effective gluons, i.e. their $\kappa$ value, is quantised into multiples of $2\delta y_g$. Thus, we may model the parton shower by generating effective gluons at the centre of discrete tiles covering the phase-space triangle. These realisations form the basis of the ``Discrete QCD algorithm" of~\cite{Andersson:1995jv}.

Each rapidity slice can be treated independently of any other slice. Inserting Eq.~\ref{eq:dqcd-inc-rate} into the exclusive decay probability (the second term in the shower master equation, Eq.~\ref{eq:psoperator}) shows that the exclusive rate for finding an effective gluon in a fixed $y$-bin takes the straightforward form
\begin{align}
dt d\xi \frac{d\phi}{2\pi} C \frac{\alpha_s}{2\pi} \frac{2s_{ik}(t,\xi)}{s_{ij}(t,\xi)s_{jk}(t,\xi)} 
\Delta(t_n,t )
 =
\frac{d \kappa}{\kappa}
\exp\left(-\int\limits^{\kappa_{max}}_\kappa \frac{d \bar\kappa}{\bar\kappa} \right)
= \frac{d \kappa}{\kappa_{max}}
\end{align}
Thus, for a fixed $y$-bin, the $\kappa$-value of an effective gluon in Fig.\ \ref{fig:dqcd} is, to first approximation, simply given by $(\textnormal{number of tiles})^{-1}$. This is significantly simpler than in conventional showers, which rely on sampling the no-emission probability $\Delta$ with the ``veto algorithm".

Once an effective gluon has been selected, new triangles are folded out of the parent region. Effective gluon positions in this fold are again quantised into tiles of dimensions $\delta y_g \times 2\delta y_g$, as illustrated by the upper left part of Fig.~\ref{fig:dqcd-baseline}. However, a simpler interpretation is possible since the height and the $y-$range of each fold are redundant: All information necessary to calculate momentum invariants $s_{ij}$ can be read off the baseline of the folded triangle structure, shown in the lower left part of Fig.~\ref{fig:dqcd-baseline}. We will call a specific baseline structure a ``grove". The shortest distance (along the baseline) between two ``tips" $i$ and $j$ can be shown to equal $\ln(s_{ij}/\Lambda^2)$. Together with the knowledge of the overall centre-of-mass energy and uniformly sampled azimuthal decay angles $\phi$, this information is sufficient to construct post-decay kinematics. 

The Discrete QCD algorithm allows a simple method to produce groves with correct rates. However, there are many ways to create the grove structures apart from the Discrete QCD algorithm. One example is shown in the lower right part of Fig.~\ref{fig:dqcd-baseline}. Since the grove structure is contained in a triangular region smaller than the original background triangle, and points on the baseline are separated from their nearest neighbours by $\delta y_g$, a two-dimensional fixed step-length random walk may be used to produce the grove structures. However, such a two-dimensional random walk is heavily constrained since it cannot extend above arbitrarily high or small $\kappa$ values, and the $y$-value of each step cannot decrease. These constraints suggest that the most straightforward algorithm to create the grove structure with correct probabilities is a one-dimensional random walk with variable step length. The step length would be determined by an auxiliary $y$-bin dependent choice, which considers that small $|y|$ values allow for fewer $\kappa$ values. An example of a one-dimensional random walk is given in the upper right part of Fig.~\ref{fig:dqcd-baseline}.

The random-walk algorithm can be summarised by\\
\vskip 0.0ex
\begin{algorithm}[H]
\DontPrintSemicolon
Start at the leftmost point on the baseline. Mark the position as quark.\\
Move to the next neighbour along the baseline with unit probability.\\
\While{walker has not reached the rightmost point on the primary baseline}{
\vskip 1ex
Calculate the number of tiles $n_{max}$ in this rapidity slice. Choose a new baseline\\
height $h=n \delta y_g$, with $n\in \{0,1,...,n_{max}-1\}$ and probabilities $P_n=1/n_{max}$.\\
\eIf{$h=0$}{
   move to the next neighbour along on current baseline with unit probability.\;
} {
  ``Step into the fold", i.e.\ take one step of length $\delta y_g$ along a new baseline\footnotemark[1] .\;
  \uIf{the position after the step $=h$} {
    Mark the position as effective gluon\footnotemark[2]\; Add one step along the current baseline with unit probability.
  } \uElseIf{the position after the step $=2h$} {
    ``Step out of the fold"\footnotemark[3]\, and move to next neighbour on the previous\; baseline with unit probability.
  }
}
}
Mark the rightmost point as antiquark.
\end{algorithm}%
\footnotetext[1]{Indicated by changing the ``line colour" in upper right part of Fig.~\ref{fig:dqcd-baseline}\,.}
\footnotetext[2]{The position $h$ marks the ``tips" of the gluon fold in the upper left part of Fig.~\ref{fig:dqcd-baseline}\,. Positions $<h$ correspond to one side of the gluon fold, positions $>h$ to the other side of the fold.}
\footnotetext[3]{Indicated by reverting to the previous colour in upper right part of Fig.~\ref{fig:dqcd-baseline}\,.}
\vskip 2.0ex
\noindent
This very compact algorithm is inspired by considering the constraints of a quantum realisation on current NISQ devices. The moderate quantum volume of these devices favours a less resource-intensive one-dimensional approach. This compact classical parton shower algorithm is an essential finding of this article and shows potential synergies between classical and quantum algorithm development.

Note that all possible groves can be enumerated for fixed centre-of-mass energy since the $(y,\kappa)$ plane is discretised in terms of $\delta y_g$. This allows pre-computing all grove structures and their rates (e.g.\ with the random walk algorithm above) before simulating event data from the information stored in the grove. Event generation then proceeds by picking a grove structure according to its probability and setting up the particle momentum vectors defined by the grove. Such an algorithm is straightforward and efficient. Moreover, during the event generation step, it is unnecessary to know ``how" the groves and their rates have been produced. In fact, we can use an elegant quantum algorithm to achieve grove generation.

\subsection{Generating scattering events from the grove structure}\label{sec:kinematics}

Event data can be generated once a grove structure has been selected. Within the classical algorithm we use as a baseline for the quantum algorithm below, we create post-decay momenta iteratively from the grove structure, creating the highest-$\kappa$ effective gluons first. For each effective gluon $j$ that has been emitted from a dipole $\mathrm{IK}$, we read off the values of $s_{ij}$, $s_{jk}$ and $s_\mathrm{IK}$ from the grove, generate a uniformly distributed azimuthal decay angle $\phi$, and then employ the momentum mapping of~\cite{Gehrmann-DeRidder:2011gkt} to produce post-branching momenta\footnote[4]{Other momentum mappings would, of course, be possible. Our choice reflects that our dipole-shower algorithm does not distinguish individual partons, leading us to believe that a momentum mapping that does not differentiate individual partons is suitable}.

There is some degree of freedom when reading off $s_{ij}$ and $s_{jk}$ from the grove structure since the latter only defines which $(y,\kappa)$-tile (in Fig.~\ref{fig:dqcd}) is picked. In fact, as argued below Eq.~\ref{eq:dqcd-inc-rate}, each point within the $(y,\kappa)$-tile is equally likely. This freedom can be used to mitigate the most glaring effects of phase-space discretisation. We follow the strategy of~\cite{Andersson:1995jv} and distribute the $y$ and $\kappa$-values of the effective gluon in the highest and second-highest $\kappa$ tiles uniformly within the tile. This ensures a good model for the highest-$\kappa$ gluons.

The simple grove selection probabilities rely on the fact that the $(y,\kappa)$ plane is uniformly covered, by virtue of Eq.\ \ref{eq:probspace}, making the selection probabilities proportional to the area of the $(y,\kappa)$-tiles. If a tile protrudes beyond the allowed phase space boundaries, its area is reduced~\cite{Andersson:1995jv}. We incorporate this correction by weighting each event with a factor 
\begin{equation}
w = \prod_{i_s}\frac{ (\textnormal{area of tile picked in slice $i_s$}) \, /\,  (\textnormal{area of slice $i_s$}) }{(\textnormal{number of tiles in slice $i_s$})^{-1}}~.
\end{equation}

\section{The Quantum Parton Shower Algorithm}\label{sec:quantumAlgo}

This section outlines the generation of the grove structures for the parton shower described in Sec.~\ref{sec:classicalAlgo} using a quantum device. The quantum algorithm has been constructed using the quantum walk framework~\cite{PhysRevA.48.1687, QWProc, Kempe}. The quantum analogue of the classical random walk, the discrete-time quantum walk defines the movement of a \textit{walker} in position space, $\mathcal{H}_P$, dictated by the result of a coin operation, $\mathcal{H}_C$, such that the total Hilbert space of the quantum walk is 

\begin{equation}
\mathcal{H} = \mathcal{H}_P \otimes \mathcal{H}_C.
\end{equation}
A single step in the quantum walk is constructed from two components: the \textit{coin flip} operation, $C$, which encodes the coin state in $\mathcal{H}_C$ onto a coin register; and the \textit{shift} operation, $S$, which controls from the coin register and moves the walker in $\mathcal{H}_P$ accordingly. The evolution of the walk is entirely unitary and can thus be expressed as 

\begin{equation}
U = S \cdot (C \otimes \mathbb{I}),
\end{equation}
where $U$ is a unitary matrix that describes a single step of the walk in $\mathcal{H}$, such that an $N$-step quantum walk is defined by $U^N$. The quantum walk has drastically different characteristics from a classical random walk due to possible interference effects introduced in the coin operation. A thorough review of the quantum walk framework is available in Reference~\cite{Kempe} and references therein. 

Quantum walks are compact algorithms which have the potential to reduce the Quantum Volume required on the device~\cite{Williams:2021lvr}. Consequently, the quantum walk architecture is well suited to designing algorithms for NISQ devices. Such quantum devices have a small number of controllable qubits and require shallow circuit depths to limit decoherence effects on the device. Appendix~\ref{app:streamlined} discusses how algorithms can be streamlined to get practical results from NISQ devices, highlighting some difficulties associated with such devices. 

Of particular interest to a quantum algorithm for parton shower simulation is the quantum walk with memory~\cite{McGettrick2010, Shakeel2014}. Quantum walks with memory can simulate arbitrary dynamics by modifying the movement of the walker based on the memory of the previous position~\cite{Camilleri2014} and coin operations~\cite{PhysRevA.67.052317, PhysRevA.87.052302}. The amount of memory that the quantum walk has affects the diffusive characteristics of the quantum walk: from an ideal quantum distribution in the limit of minimal memory to an ideal classical distribution in the limit of full memory~\cite{PhysRevA.67.052317}. Importantly, no decoherence effects are present within the algorithm, and the evolution is entirely unitary, even within the limit of full memory. Moreover, by reversing time, one can regain the initial state of the quantum walk, which is impossible from a genuinely classical algorithm~\cite{PhysRevA.87.052302}. Therefore, quantum walks with memory provide the opportunity to simulate arbitrary dynamics, including classical statistics, on a quantum device. For this reason, quantum walks with memory are of great interest in computing~\cite{PhysRevA.93.042323, Roget2020} and high energy physics~\cite{Williams:2021lvr}.

In contexts outside of high energy physics, quantum walks have been shown to exhibit speed-up over their classical counterparts for certain cases. Simulated annealing~\cite{PhysRevLett.101.130504, PhysRevA.78.042336} and search algorithms~\cite{PhysRevA.67.052307, Szegedy} are prominent examples, and it is hoped that quantum walks will provide a speed-up for generic Markov Chain Monte Carlo (MCMC) algorithms~\cite{Montanaro}. Improved speeds have been shown for dedicated MCMC algorithms under specific conditions~\cite{Montanaro, Lemieux2020efficientquantum}. The critical characteristic that leads to the speed of an MCMC algorithm is the Markov Chain mixing time, the time it takes for the MCMC algorithm to reach the equilibrium distribution~\cite{Levin2008}. Quantum algorithms can result in a quadratic speed-up over classical analogues in reaching equilibrium~\cite{PhysRevA.78.042336, PhysRevA.76.042306}, though proving that quantum walks lead to speed gains for MCMC algorithms in general, is still an active field of research~\cite{PhysRevA.76.042306, Orsucci2018fasterquantummixing,PhysRevA.104.032215}. Parton shower algorithms are related to, but due to the state memory implied by shower ordering conditions distinct from, MCMC algorithms. This makes developing a quantum walk parton shower also interesting more abstractly: as an MCMC-related class of algorithms that could benefit from quadratic speed-up. This article does not consider an assessment of potential speed-ups, as we focus on synthesising realistic data as a use case.

\subsection{Implementation on a quantum device}\label{sec:quantumImplementation}

\begin{figure}
\centering
\includegraphics[width=0.65\textwidth]{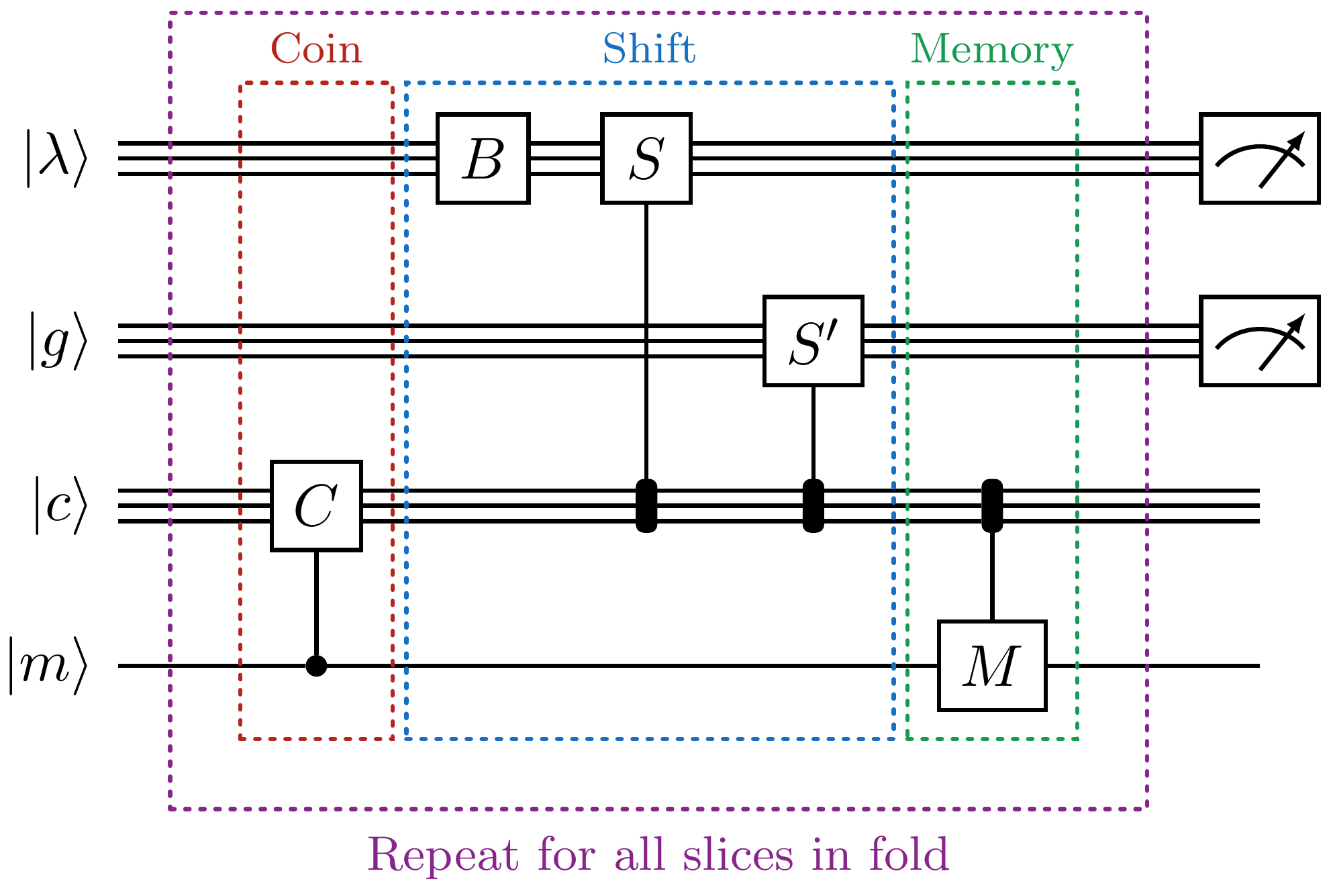}{}
\caption{\label{fig:circuitDiagram} Schematic of the quantum Discrete QCD parton shower algorithm circuit. The algorithm is a quantum walk with memory, constructed from maximum five operations per step: the coin operation $C$, baseline shift $B$, the $\lambda$ shift $S$, the gluon shift $S^\prime$, and the memory operation $M$. }
\end{figure}

With their ability to simulate arbitrary dynamics and interference effects whilst also having the potential to speed up MCMC algorithms, quantum walks are an exciting test bed for developing quantum algorithms for the simulation of parton showers. Here, we present a new approach to simulating a parton shower on a quantum device, obtaining results comparable to data from collider experiments. The quantum algorithm simulates the classical parton shower outlined in Sec.~\ref{sec:classicalAlgo}. The algorithm is constructed from a quantum walk with one dimension of memory (i.e. the memory register is constructed from one memory qubit), where the Hilbert space of the walker is formed from the grove-baseline ($\lambda$) space, $\mathcal{H}_\lambda$, defined in Fig.~\ref{fig:dqcd-baseline}, and the effective gluon space, $\mathcal{H}_g$, augmented by the coin space, $\mathcal{H}_C$. A schematic of the quantum circuit is shown in Fig.~\ref{fig:circuitDiagram}. The algorithm is constructed from a maximum of five operations per step: the coin flip operation $C$, the baseline operation $B$, the $\lambda$-shift operation $S$, the gluon-shift operation $S^\prime$, and the memory operation $M$. 

The movement of the walker in the $\lambda$-space is schematically shown in the bottom right of Fig.~\ref{fig:dqcd-baseline}. Following the algorithm's circuit shown in Fig.~\ref{fig:circuitDiagram}, the walker starts in the left-most position of the initial fold. The coin-flip operation $C$ constructs an equal state on the coin register, such that the outcome of the coin gives an equal probability of selecting a tile in the first slice, see App.~\ref{app:streamlined}. As the walker will always move to the right every step of the walk, the baseline operation $B$ is applied, irrespective of the coin outcome, to increment the walker's position in $\mathcal{H}_\lambda$. The shift operations, $S$ and $S^\prime$, then control from the coin register to move the walker correctly in $\mathcal{H}_\lambda$ and $\mathcal{H}_g$ respectively, depending on the outcome of the coin. For example, suppose the selected tile is at height $h=n\delta y_n$. In that case, the walker is moved $2n$ steps in $\mathcal{H}_\lambda$, corresponding to the walker moving along the baseline of the secondary fold created at $h$ (moving vertically down and then up in Fig.~\ref{fig:dqcd-baseline}). The walker is then shifted in $\mathcal{H}_g$ to simulate the emission of an effective gluon correctly. The final operation of the step is the memory operation $M$. This allows for a conditional coin to be applied to the register depending on the outcome of the previous coin operation, equivalent to recycling the coin. The memory register is constructed from a single qubit; thus, the walker only has a memory of the previous coin operation. Furthermore, the memory operation is only needed once for the simulation of events with $E_\mathrm{CM}=91.2~\mathrm{GeV}$, such as events from LEP experiments, thus retaining quantum diffusive characteristics. The algorithm is then repeated for any additional folds created, omitting the baseline operation, which is accounted for in the parent fold.

At the end of the algorithm, the gluon and $\lambda$ registers can be measured to obtain a grove primitive. This can then be passed to the event generation algorithm described in Sec.~\ref{sec:kinematics} to produce pseudodata from the grove structure information returned from the quantum device. It should be noted that, for exceptional cases, such as the example discussed here, the algorithm can be streamlined by removing the $\mathcal{H}_\lambda$ space. Tailoring the algorithm to simulate data from LEP experiments reduces the algorithm's extendability but decreases the circuit depth. Therefore, the algorithm can be run on NISQ devices with higher fidelity than the full algorithm. A detailed explanation of this modification is given in App.~\ref{app:streamlined}. The \texttt{ibm\_algiers} device still returns noisy results despite using the streamlined algorithm.

\subsection{Grove generation on a quantum device}

\begin{figure}[tbp]
\centering
  \begin{subfigure}{0.85\textwidth}
  \includegraphics[width=1.0\textwidth]{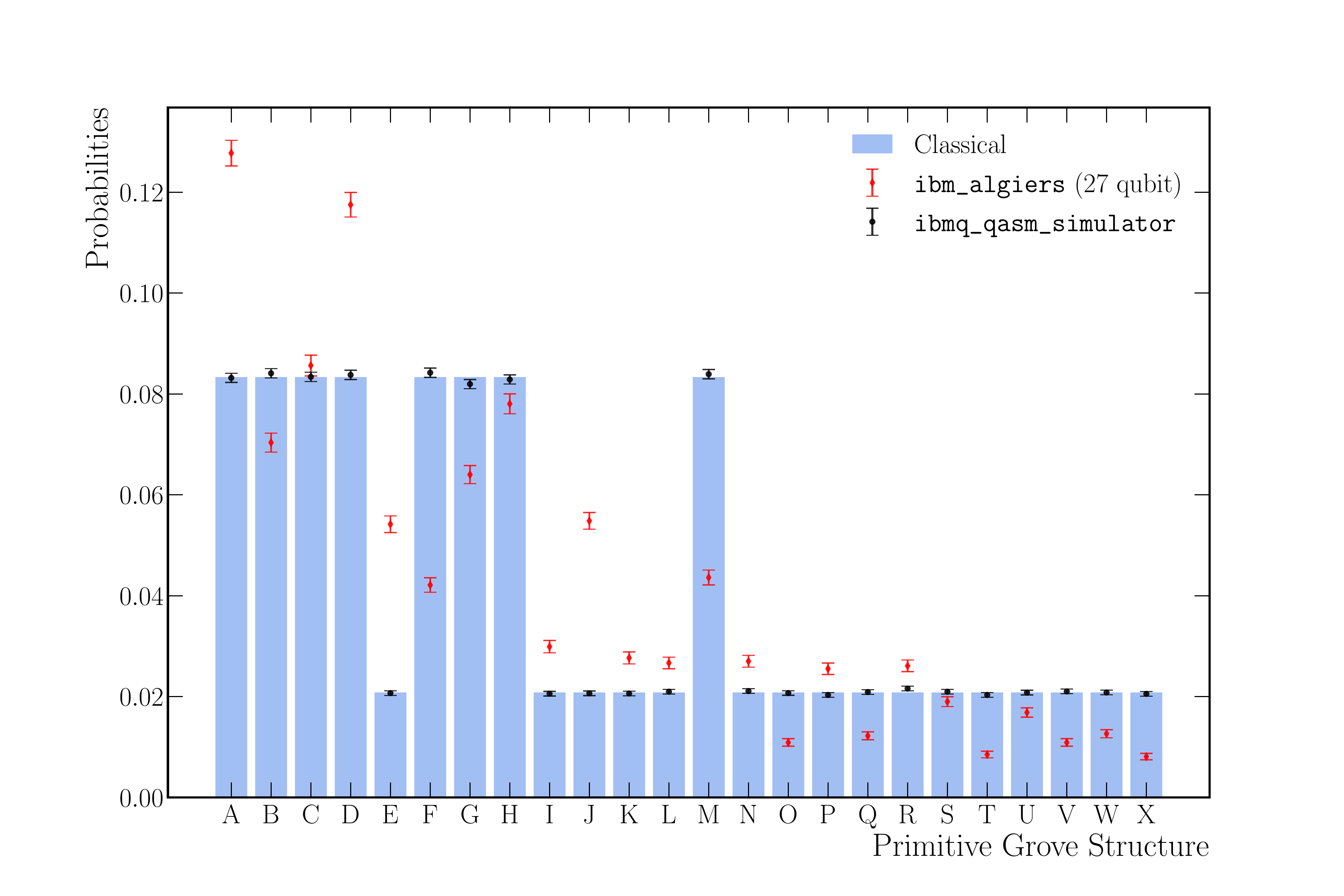}{}
  \caption{Probabilities of grove structures.}
  \end{subfigure}
  \vskip 5mm
  \begin{subfigure}{0.82\textwidth}
  \includegraphics[width=1.0\textwidth]{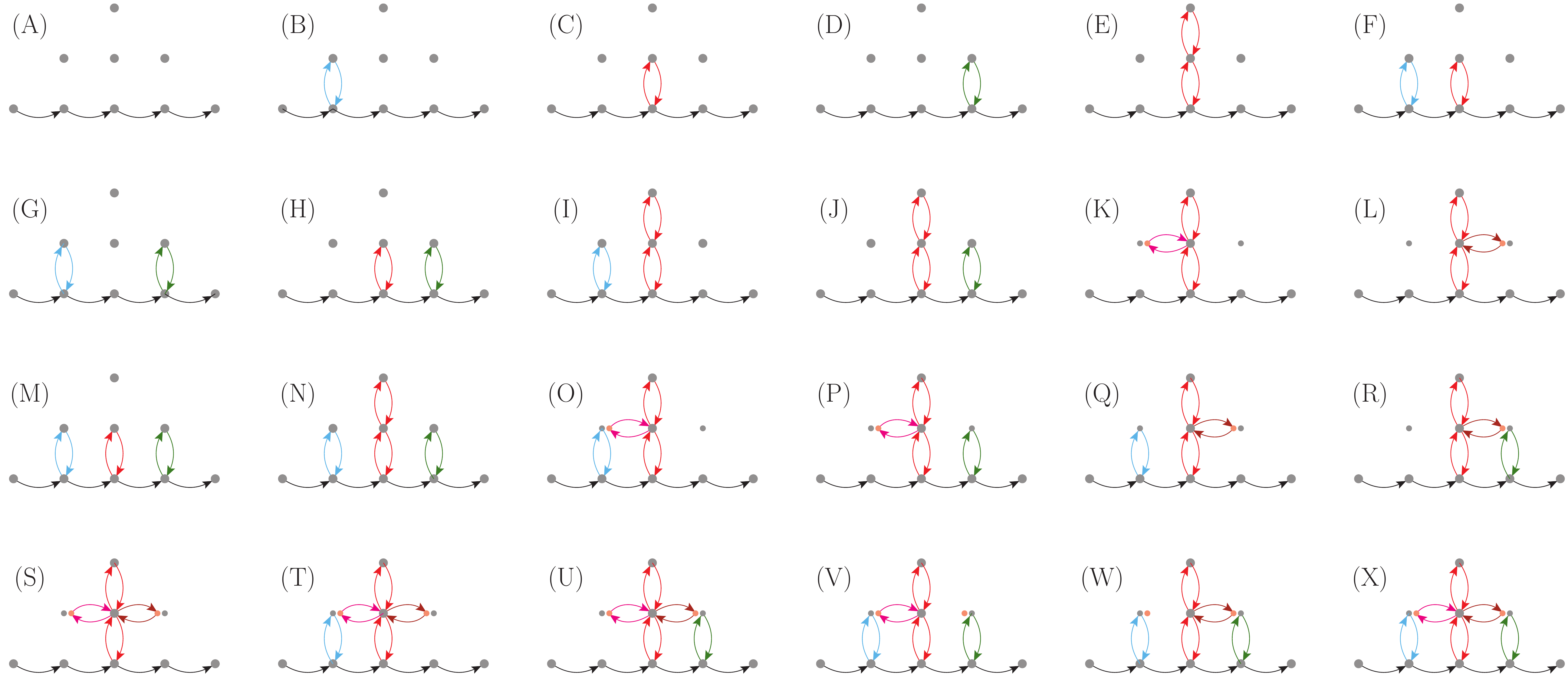}{}
  \caption{Labelling of primitive grove structures.}
  \end{subfigure}
  \caption{\label{fig:comparison} Generation of primitive grove structures for $E_\mathrm{CM}=91.2~\mathrm{GeV}$ from the 32-qubit \texttt{ibmq\_qasm\_simulator} and the 27-qubit \texttt{ibm\_algiers} device, compared to the analytically calculated classical distribution.}
\end{figure}

The parton shower algorithm presented in Sec.~\ref{sec:quantumImplementation} is the first of its kind to be comparable to real, archival collider data. To demonstrate this, the algorithm has been constructed to simulate the evolution of a $q\overline{q}$ final state produced at $E_\mathrm{CM}=91.2~\mathrm{GeV}$, typical of events at the LEP collider. For this centre-of-mass energy, there are a total of 24 primitive grove structures which can be generated, with tertiary folds being the maximum fold depth. The circuit consists of four registers, as shown in Fig.~\ref{fig:circuitDiagram}, and requires 15 qubits. The circuit is compact, using 116 gate operations (102 multi-qubit and 14 single qubit operations), and thus is suited to NISQ devices.

\begin{figure}
\includegraphics[width=0.475\textwidth]{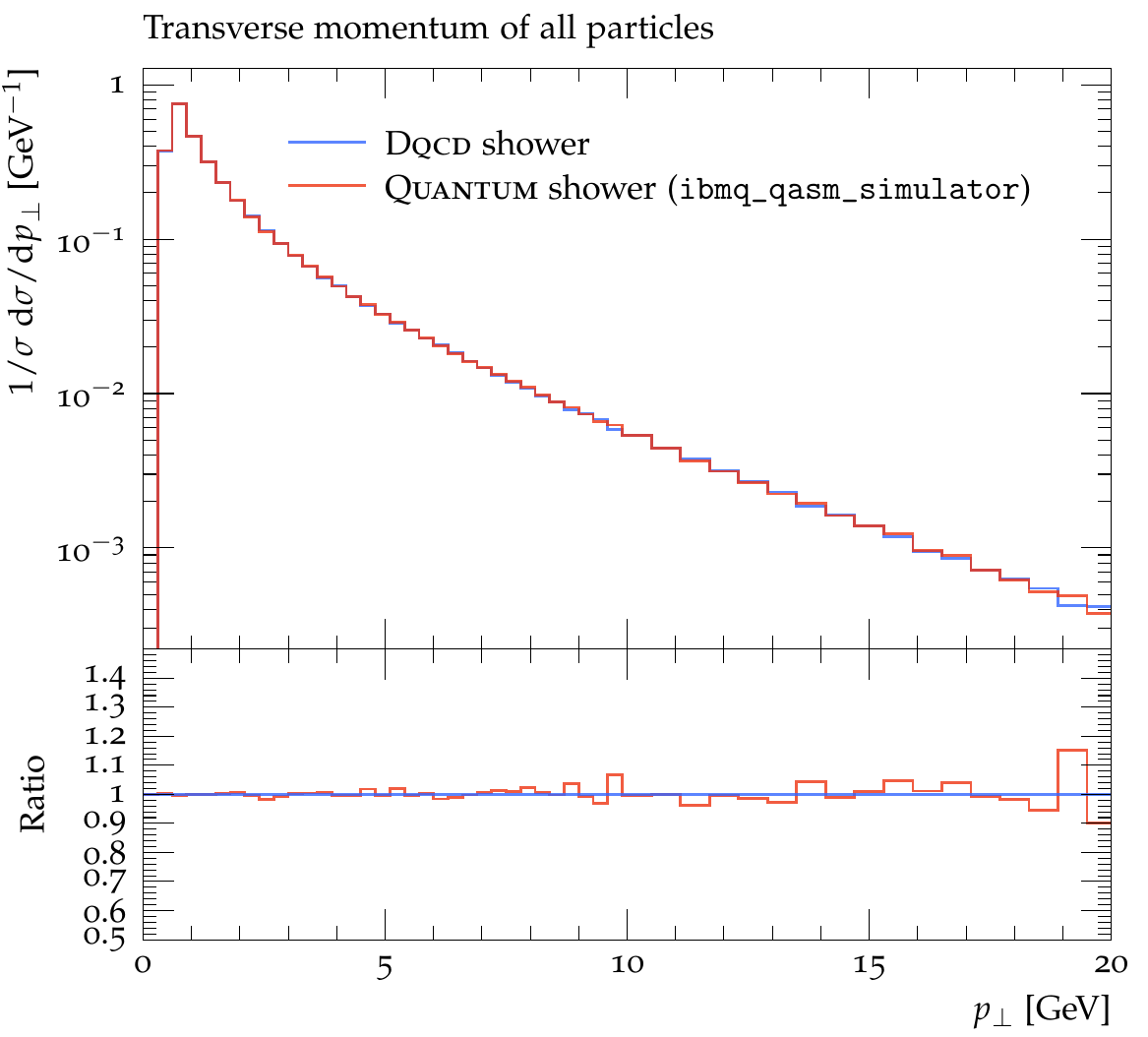}{}
\includegraphics[width=0.475\textwidth]{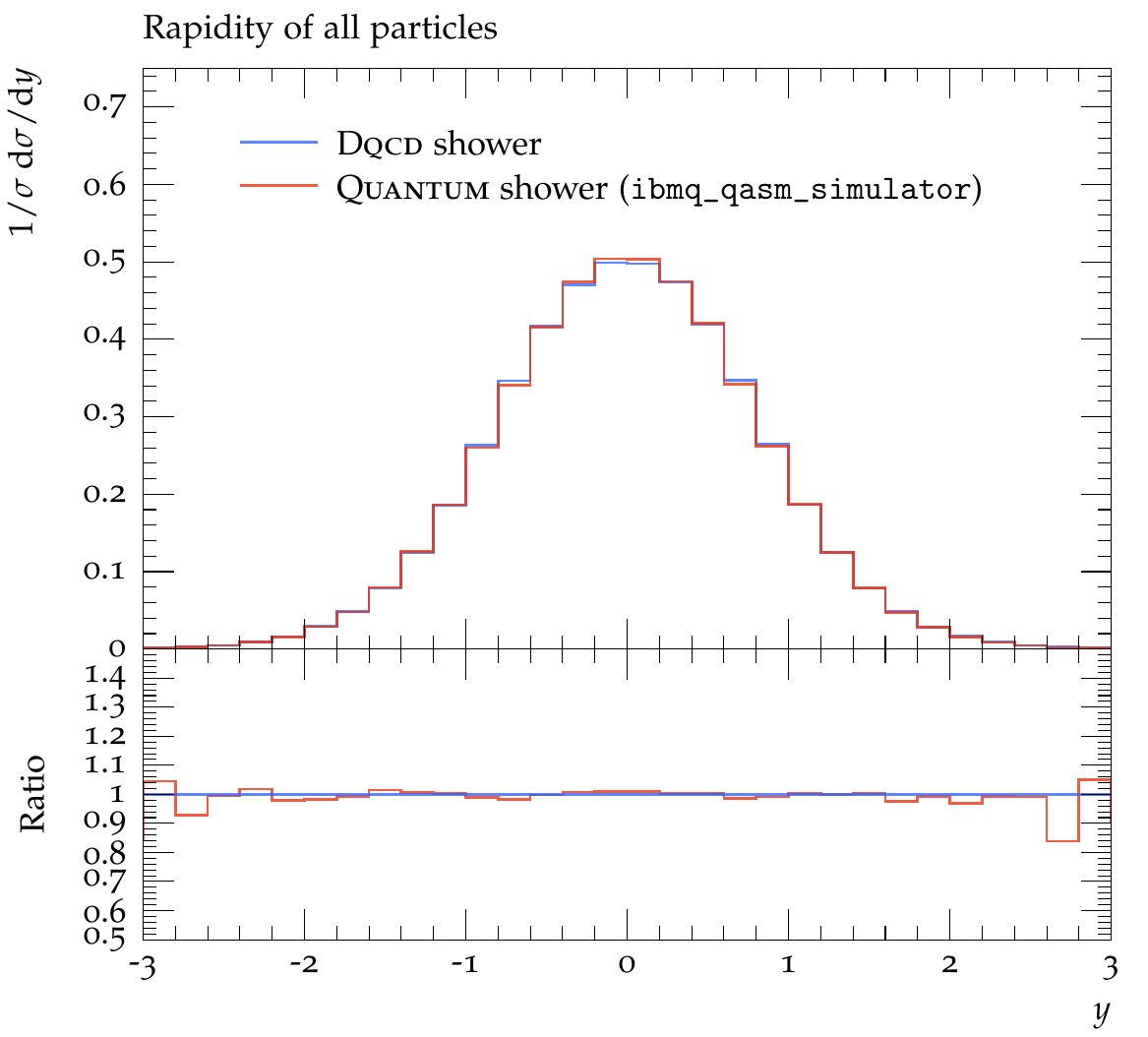}{}
\caption{\label{fig:classical-simulator-comparison} Representative comparison the (classical) Discrete QCD model and Quantum Parton Shower results from the 32-qubit \texttt{ibmq\_qasm\_simulator}, for the transverse momentum and rapidity of all final-state particles in the synthetic data. As expected from Figure \ref{fig:comparison}, the two algorithms match exactly.}
\end{figure}

Figure~\ref{fig:comparison} shows the probability distribution for grove generation from the 32-qubit \texttt{ibmq\_qasm\_simulator} and the 27-qubit \texttt{ibm\_algiers} quantum computer, compared to the analytically calculated classical distribution. The simulator has been run for 100,000 shots and shows good agreement with the classical distribution. The \texttt{ibmq\_qasm\_simulator} simulates a fully fault tolerant device and thus does not simulate noise effects or readout errors in NISQ devices. Consequently, we get an exact agreement with the classical algorithm by feeding the grove structures created by the simulator into the event generation step outlined in Sec.~\ref{sec:eventGen}. This comparison is shown in Fig.~\ref{fig:classical-simulator-comparison}.

The \texttt{ibm\_algiers} quantum computer has been run for 20,000 shots, using the tailored circuit outlined in Sec.~\ref{sec:quantumImplementation} and App.~\ref{app:streamlined}. This reduces the number of qubits in the circuit to 10 qubits and reduces the required number of gate operations from 116 gate operations to 21 gate operations (12 multi-qubit and nine single qubit gate operations). This dramatic decrease in circuit depth is crucial for obtaining high-fidelity results from NISQ devices. Figure~\ref{fig:comparison} shows the uncorrected performance of the \texttt{ibm\_algiers} device compared to the \texttt{ibmq\_qasm\_simulator} for generating the 24 grove structures for $E_\mathrm{CM}=91.2~\mathrm{GeV}$. The \texttt{ibm\_algiers} device returns noisy results, which, for some grove structures, greatly differ from the simulator. This algorithm's main source of error is a large amount of \textsc{swap} operations needed to correctly implement the gluon shift, resulting in many multi-qubit gate operations. Multi-qubit gates, such as the \textsc{cnot} gate, can introduce environmental noise to the device, resulting in decoherence.
Consequently, the many \textsc{swap} operations needed in the computation directly affect the fidelity of the results. Furthermore, the results tend to prefer states with fewer effective gluons. This is likely due to the relaxation time of the qubits in the gluon register, causing the qubits to return to the ground state before measurement. A detailed discussion of the noise effects and how these affect the results from the \texttt{ibm\_algiers} device is given in App.~\ref{app:noise}. 

It is possible to employ quantum error mitigation schemes to suppress readout errors in the results. However, it is shown in Sec.~\ref{sec:eventGen} that mitigation schemes are not needed for event generation, and the algorithm is remarkably robust to noise effects.

\section{Event Generation using the Quantum algorithm}\label{sec:eventGen}

In this section, we compare the result of using the {\tt ibm\_algiers} device to create data and compare the measurements of selected kinematic distributions performed on these simulated data with archival collider data. We aim to highlight different aspects of the simulation and thus compare both event shape and jet observables. The value of the \emph{thrust} observable $T$~\cite{ALEPH:2003obs} classifies the event into ``pencil-like" ($T\sim 1$) and ``planar-circular" events ($T\sim 0.5$). Thus, values of $T\rightarrow 1$ provide an excellent probe of the modelling of soft- and collinear parton emissions in the parton shower, and the transition between partons and hadrons for $T\sim 1$, while moderate $T$-values probe high-energy gluon emissions. The \emph{energy-energy correlation} (EEC) is another event shape observable that probes QCD evolution~\cite{DELPHI:1996sen}. Suppose the high-energy primary partons mostly retain their direction (as would be the case for the asymptotically free limit of QCD). In that case, one expects a particle at angle $\xi$ to be accompanied by another particle at angle $\xi+\pi$. As the parton shower evolution leads to collinear radiation, one expects several collimated particles around a test particle at an angle $\xi$. At intermediate EEC values from $-\frac{1}{2}\lesssim \cos\xi\lesssim \frac{1}{2}$, we expect soft-gluon radiation as well as strong-coupling (non-perturbative) QCD effects to contribute to form a plateau. The \emph{Durham 2-jet rate} $y_{23}$~\cite{JADE:1999zar} defines a ``closeness" measure of the second and third-hardest jets in the collision event\footnote[6]{At LEP, each event is expected to have at least two jets, producing two high-energy primary partons in the original $e^+e^-$-annihilation.}. Well-separated jets, meaning jets with high energy and large angular separation, lead to high $y_{23}$ values. At such values, higher-order real-emission corrections dominate the spectrum. At the intermediate 2-jet rate, the jets start to coalesce. This region is susceptible to parton shower evolution. Finally, at very small $y_{23}$ values, the jets are no longer separated, and non-perturbative (hadronisation) effects dominate the distribution. Finally, the \emph{jet mass difference} $M_D$~\cite{ALEPH:2003obs} measures the (normalised) invariant mass of particles in the left and right detector hemispheres. Small values of $M_D\sim 0$ correspond to equally populated hemispheres, as would be the case for perfectly balanced jets, while values of $M_D\lesssim \frac{1}{3}$ probe the modelling of the hard- and multiple soft-gluon emission. Large $M_D$-values probe multiple hard gluon emissions.

We embed the Quantum Parton Shower into a classical toolchain to facilitate comparisons to data from the Large Electron-Positron collider (LEP). The momentum distribution of quarks and antiquarks in the annihilation $e^+e^-\rightarrow Z/\gamma^* \rightarrow q \bar q$ at $E_\mathrm{CM}=91.2\,\mathrm{GeV}$ is generated classically. The parton shower evolution of the $q \bar q$ final state is handled by the NISQ device, which generates grove structures. We run the shower evolution algorithm on the {\tt ibm\_algiers} device, which operates a Falcon r5.11 architecture. The grove structures are then used to generate the desired kinematics, and the result is stored in Les Houches Event (LHE) File format~\cite{Alwall:2006yp,Andersen:2014efa}. These LHE files are passed to the \textsc{Pythia} event generator~\cite{Bierlich:2022pfr} for parton-to-hadron conversion by means of the Lund string model~\cite{Andersson:1983ia}. Finally, we use \textsc{Rivet} analysis framework~\cite{Buckley:2010ar,Bierlich:2019rhm} to compare our synthetic data to data taken with LEP detectors.

We do not embark on a dedicated program to ``tune" the parameters of the simulation~\cite{Buckley:2011ms} to improve the data description. The parton shower mass scale $\Lambda$ is chosen as $\Lambda=0.4$~GeV since such a value allows for a reasonably rich structure of possible groves at LEP collision energies. As mentioned in Sec.~\ref{sec:kinematics}, for the highest and second-highest $\kappa$ values we distribute the y and $\kappa$ values for the effective gluons uniformly within the tile. The parameters of the Lund symmetric fragmentation function~\cite{Andersson:1983ia} are set to $a=2$ and $b=3/5$, and width of the non-perturbative string-$p_\perp$ is set to $\sigma=0.1$~GeV\footnote[7]{A low value of $\sigma$ is motivated by the fact that the mass scale $\Lambda=0.4$~GeV also sets the shower cut-off. Thus, the shower covers lower $p_\perp$-values than conventional methods and thus requires a smaller non-perturbative component to the $p_\perp$ distribution.}.  

\begin{figure}
\includegraphics[width=0.475\textwidth]{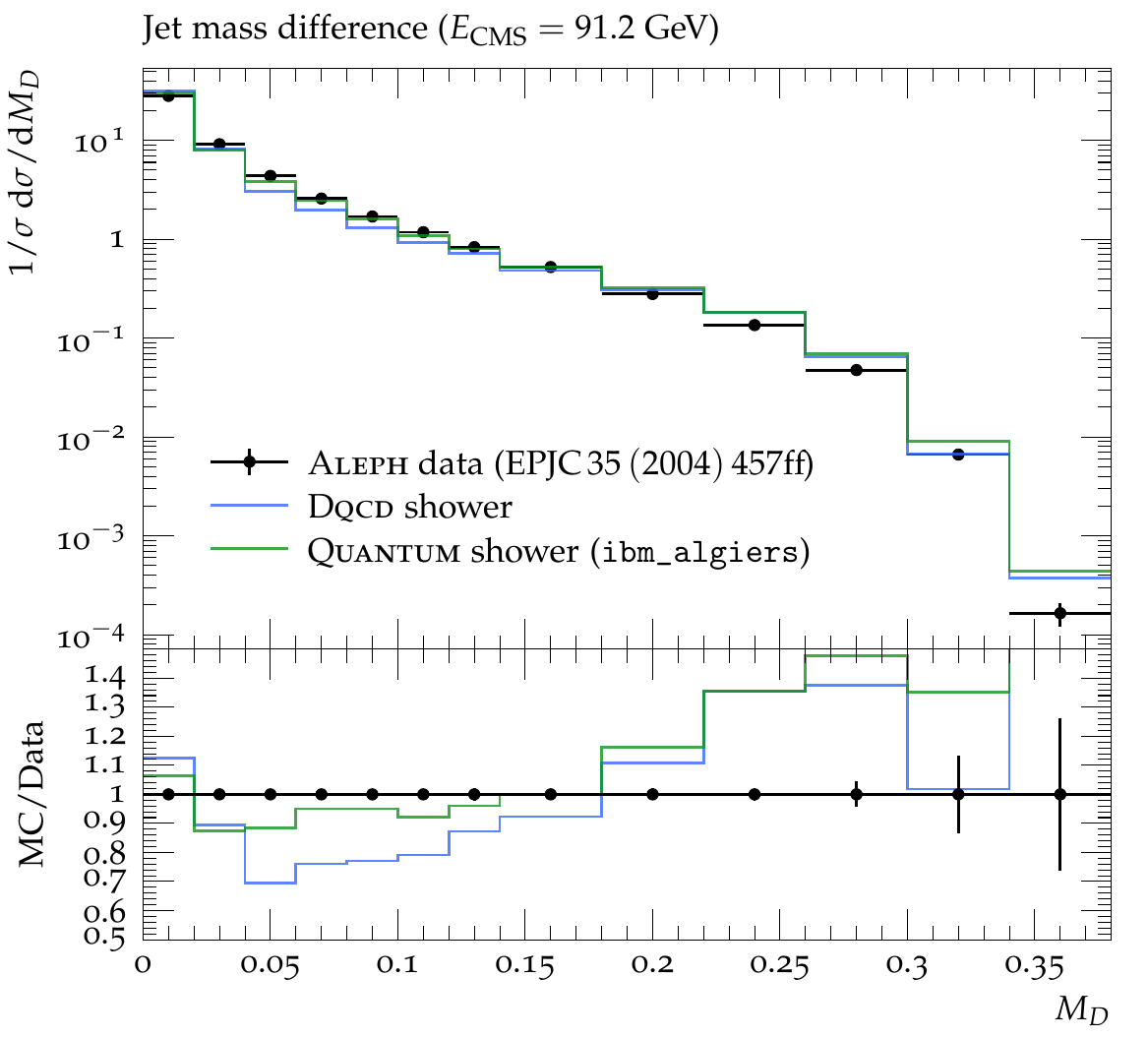}{}
\includegraphics[width=0.475\textwidth]{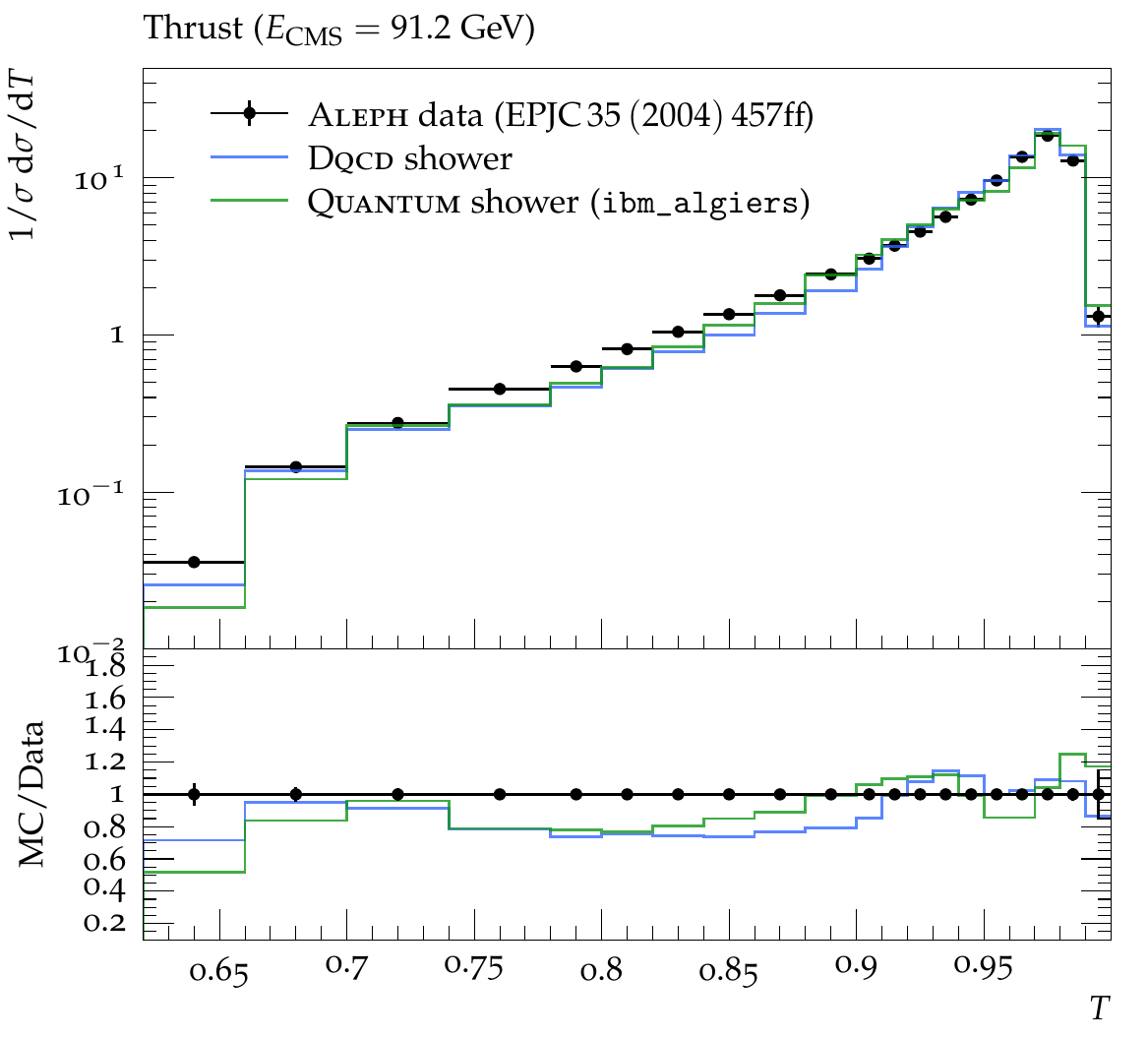}{}\\
\includegraphics[width=0.475\textwidth]{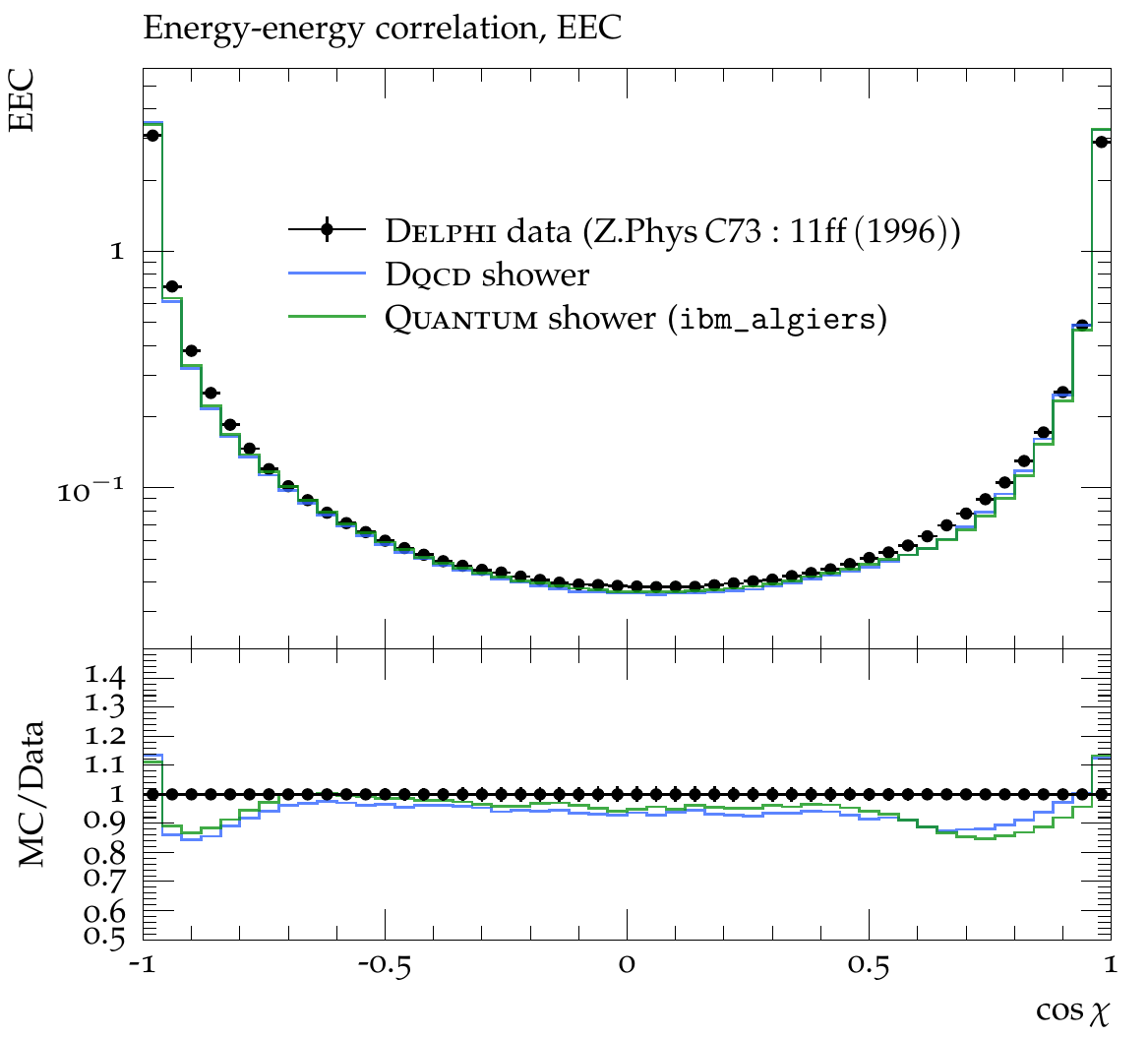}{}
\includegraphics[width=0.475\textwidth]{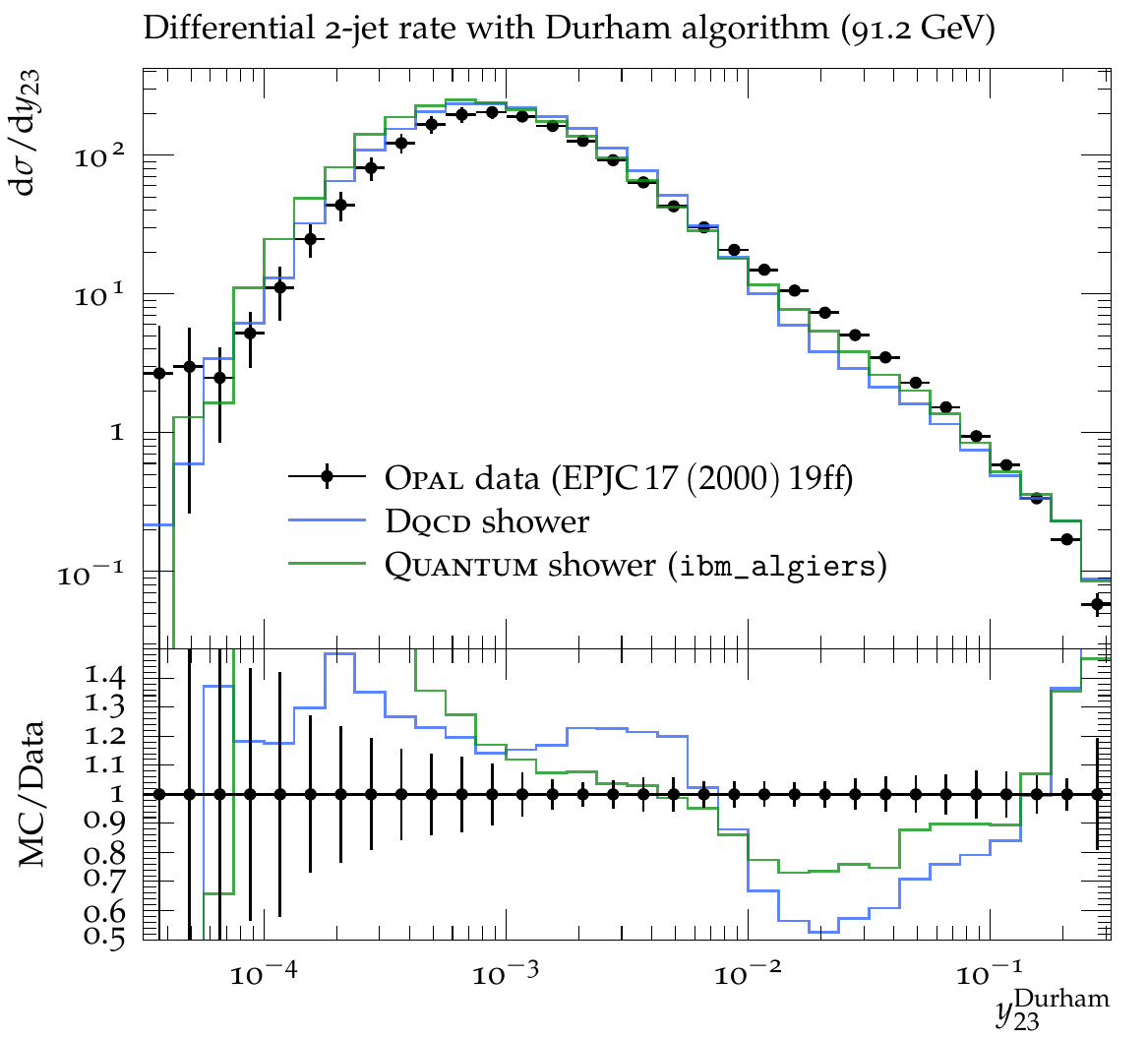}{}
\caption{\label{fig:lep-comparison} Sample comparisons of the Discrete QCD model and the Quantum Parton Shower to data taken at the LEP collider~\cite{ALEPH:2003obs,DELPHI:1996sen,JADE:1999zar}. The Quantum Parton Shower results are not corrected for errors in the qubit evaluations.}
\end{figure}

Sample comparisons to experimental data are shown in Fig.~\ref{fig:lep-comparison}. We find very satisfactory agreement of the classical \textsc{Dqcd} shower results, in particular given the simplicity of the shower model. As seen in the rightmost bins of the jet mass difference $M_D$ and the Durham 2-jet rate $y_{23}$, the algorithm tends to overpopulate states with multiple high-energy gluons. This is expected since the soft-gluon emission rate overshoots the true matrix elements in the hard region. Similarly, we expect that improving the description outside the soft limit will improve the description at intermediate thrust values $T\sim 0.75-0.9$. Both high $T$-values and moderate EEC values indicate that the emission of soft gluons is well-described.

Comparing the results of running the quantum algorithm on the {\tt ibm\_algiers} device to experimental data is again very favourable. Interestingly, the device's tendency to produce ``too few" excited (multi-gluon) states slightly improves the overall data description. Error correction would, of course, mitigate this effect. Nevertheless, the sound data description of the uncorrected quantum results may hint at improvements in the classical algorithm.

\section{Summary}

In this article, we have synthesised realistic particle collision events by sampling parton shower configurations on a quantum device. As an application, we have compared the simulated data to data recorded at experiments at the Large Electron-Positron (LEP) collider, finding favourable agreement. 

To obtain these results, we have re-interpreted conventional parton showers -- which rely on an intricate form of rejection sampling -- as one-dimensional random walks in the $\lambda$-measure~\cite{Andersson:1988ee}, i.e.\ along the baseline of the multigluon phase-space structure. The novel algorithm is a reformulation of the Discrete QCD algorithm~\cite{Andersson:1995jv}. It allows selecting the result of the parton shower evolution before multiparticle momenta are generated while ensuring on-shell conditions and momentum conservation. Developing this unique strategy was necessary to set a suitable starting point for quantum algorithm development. Furthermore, the generation and sampling of parton shower results can be performed on a quantum device, whose outputs are then used to reconstruct kinematics and ultimately perform comparisons to data.

This is the first time that the result of a quantum algorithm has been compared to ``real-life" particle physics data. The quantum algorithm is constructed using the quantum walk framework and, consequently, is a compact algorithm with a short circuit depth, an important aspect to consider to obtain practical results from NISQ devices. We synthesise data using the \texttt{ibm\_algiers} device, which uses an IBM Falcon r5.11 processor. Figure~\ref{fig:comparison} shows that the results from the \texttt{ibm\_algiers} device are subject to a non-negligible amount of noise. The main noise source has been identified as \textsc{cnot} errors due to gate decompositions and \textsc{swap} operations implemented in the transpilation process\footnote[8]{Transpilation is the process of rewriting a given input circuit to match the topology of a specific quantum device, and/or to optimise the circuit for execution on present day noisy quantum systems.}. In addition to \textsc{cnot} errors, it can be seen from the results in Fig.~\ref{fig:comparison} and \ref{fig:lep-comparison} that the algorithm prefers final states will fewer soft emissions. From a close study of the device output, this error has been attributed to the T1 (relaxation time) of the qubits in the gluon register, causing decoherence during the algorithm's run time. A detailed discussion of the errors on the device has been provided in App.~\ref{app:noise}. Cumulatively, these errors result in a device output far from ideal. However, since physical observables are supported by an admixture of several parton shower structures, our results are remarkably robust to noise, as demonstrated by the data comparison in Fig.~\ref{fig:lep-comparison}. Overall, our current quantum algorithm improves on previous quantum shower attempts in several physics aspects (the inclusion of soft-gluon coherence, running-coupling effects and realistic scattering data generation) as well as on computational aspects: the required quantum volume to simulate data from LEP experiments is less than that needed for all previously known quantum shower algorithms~\cite{Bauer:2019qxa,Bepari:2020xqi,Williams:2021lvr}. Furthermore, the necessity for fully fault tolerant devices has been drastically reduced.

A quantum event generator would also include quantum algorithms to sample the highest-momentum transfer interactions and quantum algorithms for non-perturbative hadronisation models. Progress on the former has been reported in~\cite{Bepari:2020xqi}, while~\cite{Klco:2018kyo} reported on modelling string breaking in a simplified model on a NISQ device. Combined with these advances, our results indicate that a programme of developing quantum event generators on near or intermediate time scales is within reach. This programme should also include several improvements to our quantum shower algorithm. An extension of the classical algorithm to include $g\rightarrow q\bar q$ splittings, hard-collinear corrections, and subleading colour terms may be inspired by previous work~\cite{Gustafson:1992uh}, though detailed studies are necessary. A programme of dedicated event generator parameter ``tuning", as is typical for conventional event generators, should also be initiated for the quantum shower algorithm. To obtain competetive predictions, appreciable work will be necessary to incorporate precision QCD calculations into the Discrete QCD framework through matching
~\cite{Frixione:2002ik,Nason:2004rx,Frixione:2007vw} or merging~\cite{Catani:2001cc,Lonnblad:2001iq,Mangano:2001xp,Mrenna:2003if,Alwall:2007fs} methods.

Notwithstanding these challenges, it is exciting to see that the rapid advances in algorithm design on NISQ devices allow us to think about such technicalities already today.

 

\section*{Acknowledgments}

 We thank Sarah Malik for valuable discussions. We acknowledge the use of IBM Quantum services for this work and to advanced services provided by the IBM Quantum Researchers Program. The views expressed are those of the authors, and do not reflect the official policy or position of IBM or the IBM Quantum team. We would like to specifically thank the IBM Support Centre for their help with running on the
 \texttt{ibm\_cloud}.

\appendix

\section{Tailoring of the Quantum Implementation}\label{app:streamlined}

\begin{figure}[b]
\centering
\includegraphics[scale=0.6]{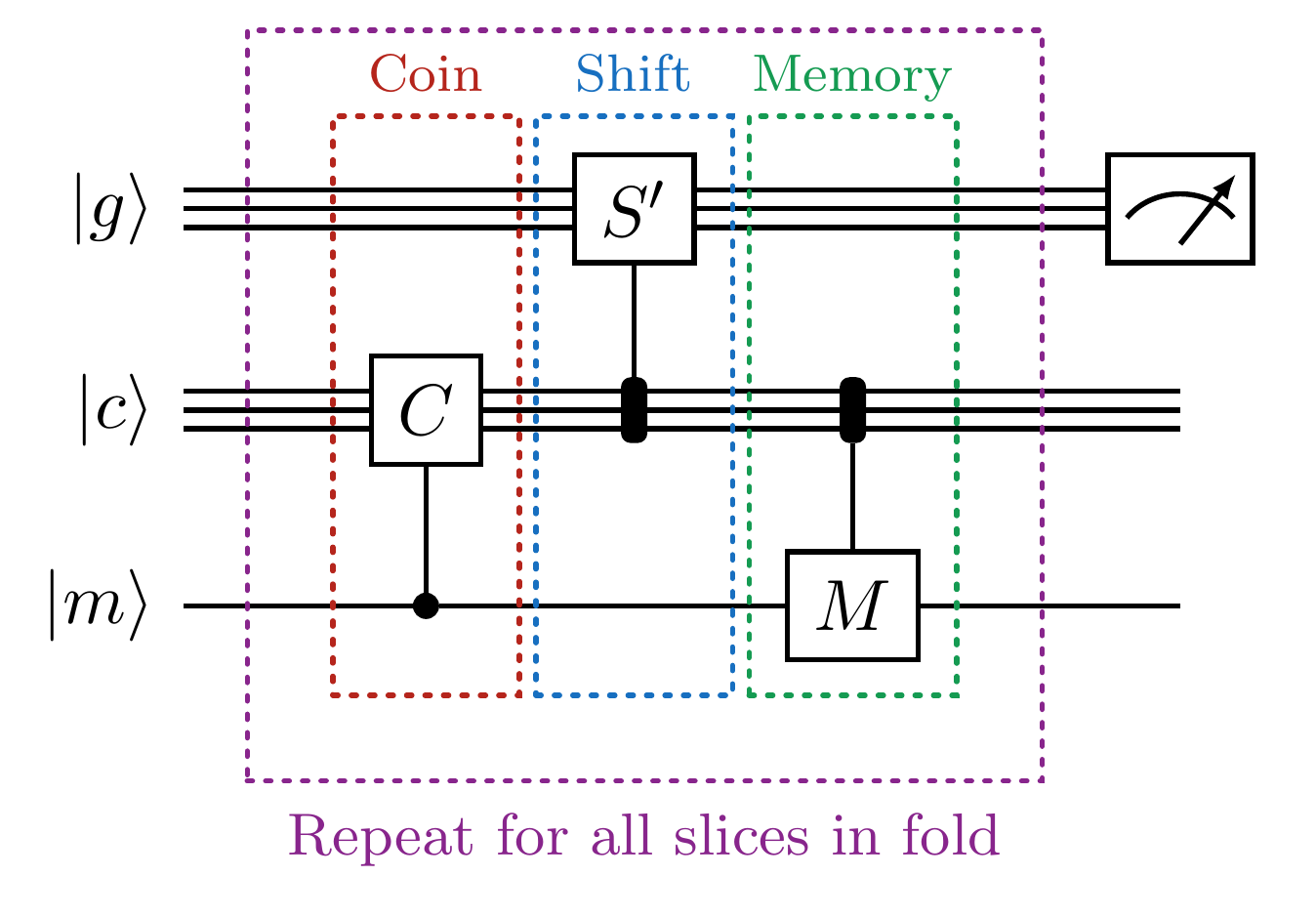}{}
\caption{\label{fig:reducedCircuit} Schematic of the reduced quantum Discrete QCD parton shower algorithm circuit. The algorithm is a quantum walk with memory, constructed from maximum three operations per step: the coin operation $C$, the gluon shift $S^\prime$, and the memory operation$M$.}
\end{figure}

Noisy Intermediate-Scale Quantum (NISQ) devices have a limited number of qubits and, perhaps more importantly, require shallow circuit depths to return high fidelity results. Furthermore, current quantum error mitigation schemes are not to a high enough standard to suppress all errors from the device, such as read-out and \textsc{cnot} errors. Therefore, to run practical algorithms on current NISQ devices, it is essential to write efficient quantum algorithms which can minimise the required quantum volume on the device, thus reducing noise throughout the calculation. 

The general and extendable quantum algorithm for the simulation of parton showers outlined in this paper is the first quantum algorithm with the ability to simulate a full parton shower with kinematics. The algorithm is implemented using the quantum walk framework, allowing for compact and efficient circuit implementation, as shown in Fig.~\ref{fig:circuitDiagram}. For a centre of mass energy typical of collisions at the LEP collider, the algorithm can be run on the 27-qubit \texttt{ibm\_algiers} device and is constructed from 15 qubits and 116 gate operations (102 multi-qubit and 14 single qubit gate operations). However, \texttt{ibm\_algiers} is a NISQ device, and even the compact circuit depth of the parton shower algorithm introduces significant noise effects and produces low fidelity results. For the specific example discussed in this paper, it is possible to streamline the algorithm at the expense of the extendability of the algorithm but at the benefit of drastically reducing the quantum volume required. 

Removing the $\lambda$-walk in $\mathcal{H}_\lambda$ from the quantum walk, restricting the walker to only $\mathcal{H}_g$ augmented by the coin space, $\mathcal{H}_C$, allows for the circuit to be reduced in size to 10 qubits and 21 gate operations (12 multi-qubit and nine single qubit gate operations). Crucially, the number of multi-qubit gate operations is decreased by a factor of approximately five, resulting in a decrease in the \textsc{cnot} errors within the device. A schematic of this reduced circuit is shown in Fig.~\ref{fig:reducedCircuit}. 

This algorithm has been run for 20,000 shots on the \texttt{ibm\_algiers} device through the \texttt{ibm\_cloud} service. The graph for the grove generation is shown in Fig.~\ref{fig:comparison}, and highlights the errors on the device compared to the output from the \texttt{ibmq\_qasm\_simulator}. The error on the device remains large, despite the use of the streamlined algorithm. Section~\ref{sec:eventGen} discusses the event generation using the distribution of primitive grove structures produced by the quantum device. Remarkably, the event generation algorithm is robust to noise on the quantum device. 

The streamlined algorithm uses a quantum walk with memory framework, discussed in Sec.~\ref{sec:quantumAlgo}. For a centre of mass energy $E_\mathrm{CM}=91.2\,\mathrm{GeV}$, the memory operation is only required once in the algorithm. This is applied after the first step to implement the correct coin operation for the second step. Consequently, the walker remains within the limit of quantum diffusive characteristics. 

\begin{figure}[h!]
\centering
\includegraphics[scale=0.6]{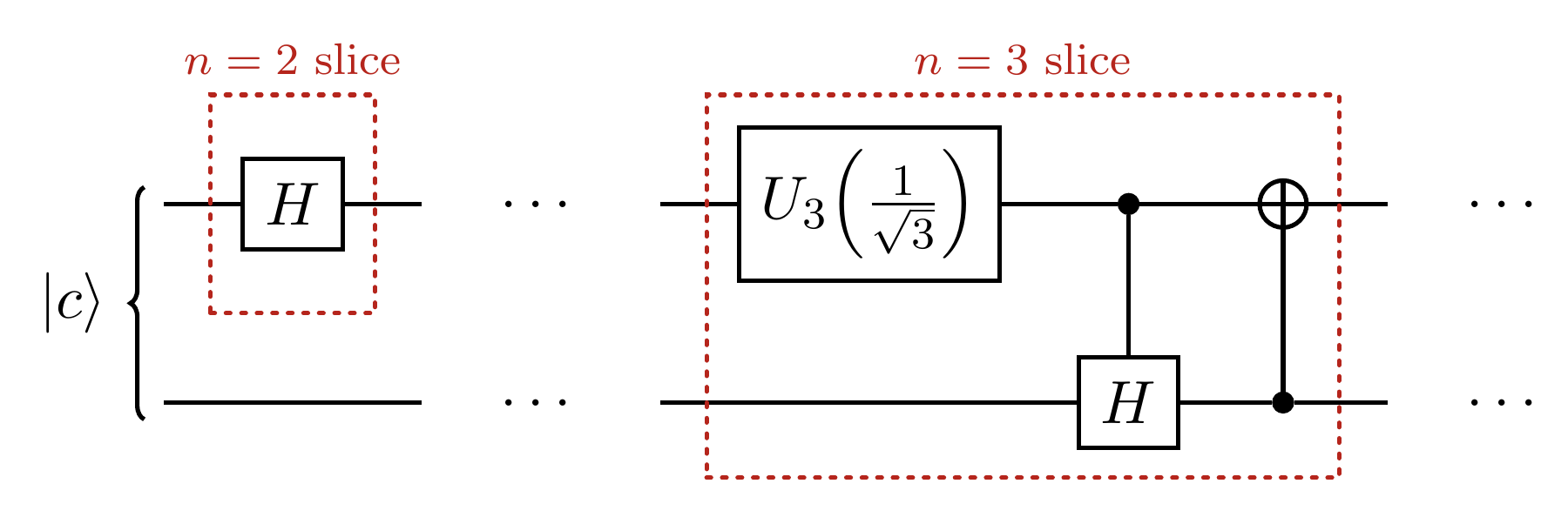}{}
\caption{\label{fig:coinOp} Coin operations for slices with $n=2$ and $n=3$ tiles. }
\end{figure}

The coin operation constructs an equal superposition of $n$ states on the coin register of $m$ qubits, such that the outcome of the coin gives an equal probability of obtaining each state, 

\begin{equation}
C\vert 0 \rangle^{\otimes m} = \frac{1}{\sqrt{n}} \Big( \vert 0 \rangle + \vert 1 \rangle + \hdots + \vert n -1 \rangle \Big).
\end{equation}
This reflects the probability of selecting a tile in a slice of $n$ tiles, see Fig.~\ref{fig:dqcd}. As an example, consider the centre of mass energy $E_\mathrm{CM}=91.2\,\mathrm{GeV}$. The phase space available is constructed from slices of two and three tiles, requiring coin register states,

\begin{align}
\frac{1}{\sqrt{2}} \Big ( \vert 0 \rangle + \vert 1\rangle \Big), &&\textrm{and} && \frac{1}{\sqrt{3}} \Big( \vert 00 \rangle + \vert 10 \rangle + \vert 01 \rangle \Big),
\end{align} 
respectively. These states can be constructed simply using the circuit diagrams shown in Fig.~\ref{fig:coinOp}.

\section{Identification of Quantum Errors}\label{app:noise}

The use of Noisy Intermediate-Scale Quantum (NISQ) devices is impractical without understanding the possible sources of quantum errors present within the device~\cite{RevModPhys.82.1155}. Unlike classical devices, quantum computers are limited in qubit number and thus cannot rely on redundancy for error correction~\cite{FrontEng}. For that reason, quantum error correction is a highly active research field~\cite{lidar2013quantum, Devitt_2013}. However, a general scheme for error correction, resulting in a fully fault tolerant device, does not currently exist. Quantum devices are subject to environmental decoherence effects, such as unwanted electric fields or hardware defects~\cite{FrontEng}. Therefore, to obtain high fidelity results from a NISQ device, careful considerations of circuit architecture have to be made, most importantly the depth of the circuit and the number of multi-qubit operations. For this reason, the quantum shower algorithm outlined in Sec.~\ref{sec:quantumImplementation} has been streamlined for the special case of $E_\mathrm{CM}=91.2\,\mathrm{GeV}$. The tailoring of this special case is outlined in App.~\ref{app:streamlined}. 

The quantum parton shower algorithm has been run on the \texttt{ibm\_algiers} devices for 20,000 shots. A comparison to analytically calculated expected probabilities and simulated quantum results from the \texttt{ibmq\_qasm\_simulator} is given in Fig.~\ref{fig:comparison}. It is clear from the disparity between the \texttt{ibm\_algiers} device and the \texttt{imbq\_qasm\_simulator} that the real quantum device returns noisy results, with the tendency to prefer grove structures with less effective gluons produced, a visible effect in Fig.~\ref{fig:comparison} and \ref{fig:lep-comparison}. The high number of multi-qubit gate operations is the main source of error for the quantum shower algorithm on the \texttt{ibm\_algiers} device. Currently, on the \texttt{ibm\_algiers} device, multi-controlled gates such as the \textsc{ccnot} gate are decomposed into \textsc{cnot} gates. Therefore, even if only a small amount of \textsc{ccnot} gates are used, including the required \textsc{swap} operations, the number of \textsc{cnot} gates implemented can be an order of magnitude higher. Consequently, this will greatly affect the results' fidelity as \textsc{cnot} gates can introduce environmental noise leading to decoherence.

Another primary source of error in the results from Fig.~\ref{fig:comparison} comes from the qubits' T1 time (sometimes called the relaxation time). The T1 time of a qubit is the time it takes the qubit to decay from the excited state, in this case, $\vert 1 \rangle$, to the ground state, $\vert 0 \rangle$ on IBM devices~\cite{QiskitT1}. The T1 time, therefore, dictates the time in which practical and meaningful algorithms can be run. For the quantum shower algorithm, the gluon shift records whether a gluon has been created by applying a shift in $\mathcal{H}_g$ to the walker, as shown in Fig.~\ref{fig:circuitDiagram}. As more gluons are created, more qubits are flipped from the $\vert 0 \rangle$ to the $\vert 1 \rangle$ state. If the first slice produces a gluon, then some qubits in the gluon register will be flipped to the $\vert 1 \rangle$ state and can remain in this state throughout the algorithm. These qubits could likely decay to the ground state $\vert 0 \rangle$ whilst the algorithm is still running on the device and therefore would lead to incorrect grove structures measured at the end of the algorithm. This would explain the excess of events without soft emission. A thorough overview of hardware noise and errors is given in~\cite{QiskitT1} and references therein.

\begin{figure}[t]
\includegraphics[width=0.475\textwidth]{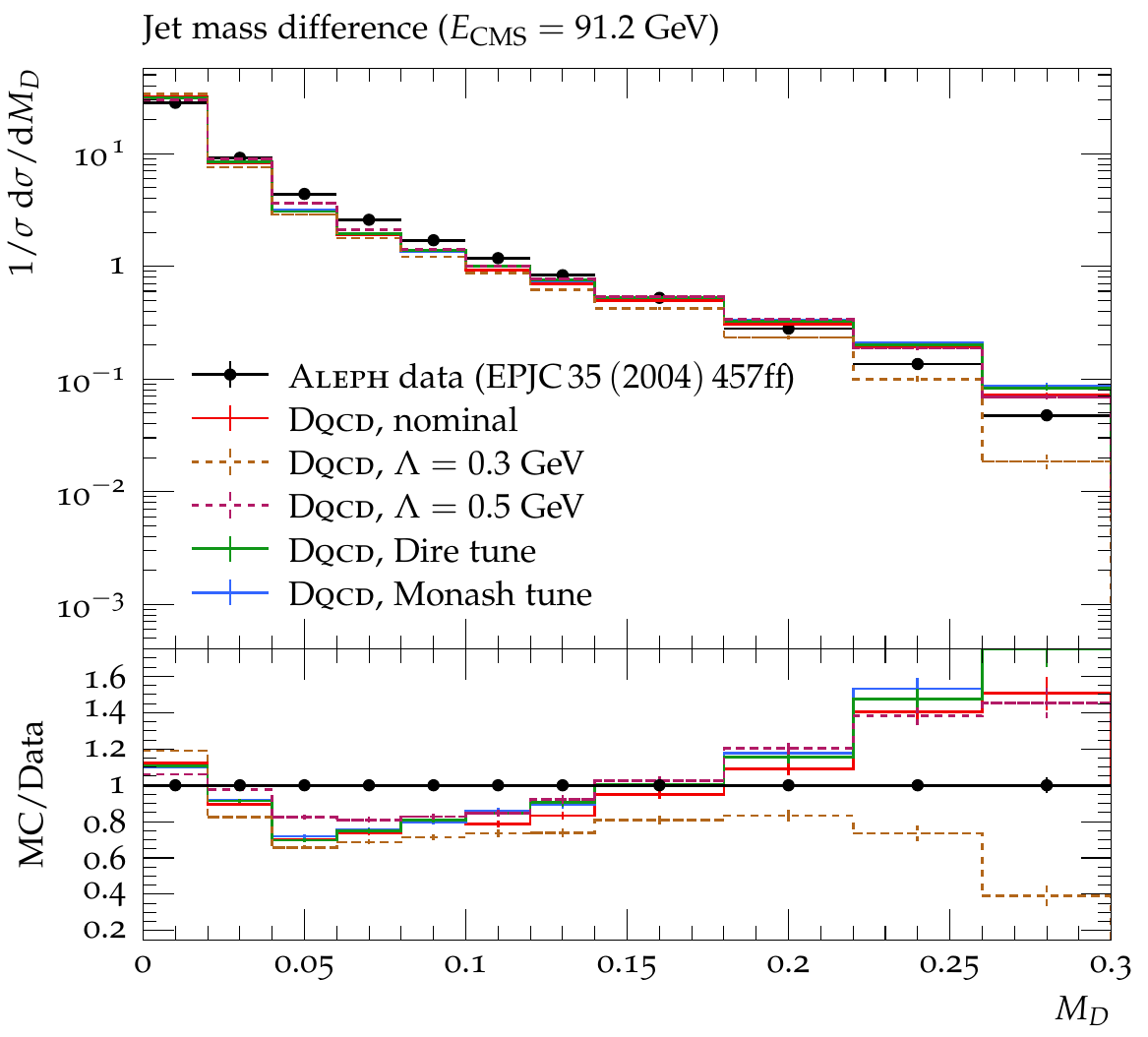}{}
\includegraphics[width=0.475\textwidth]{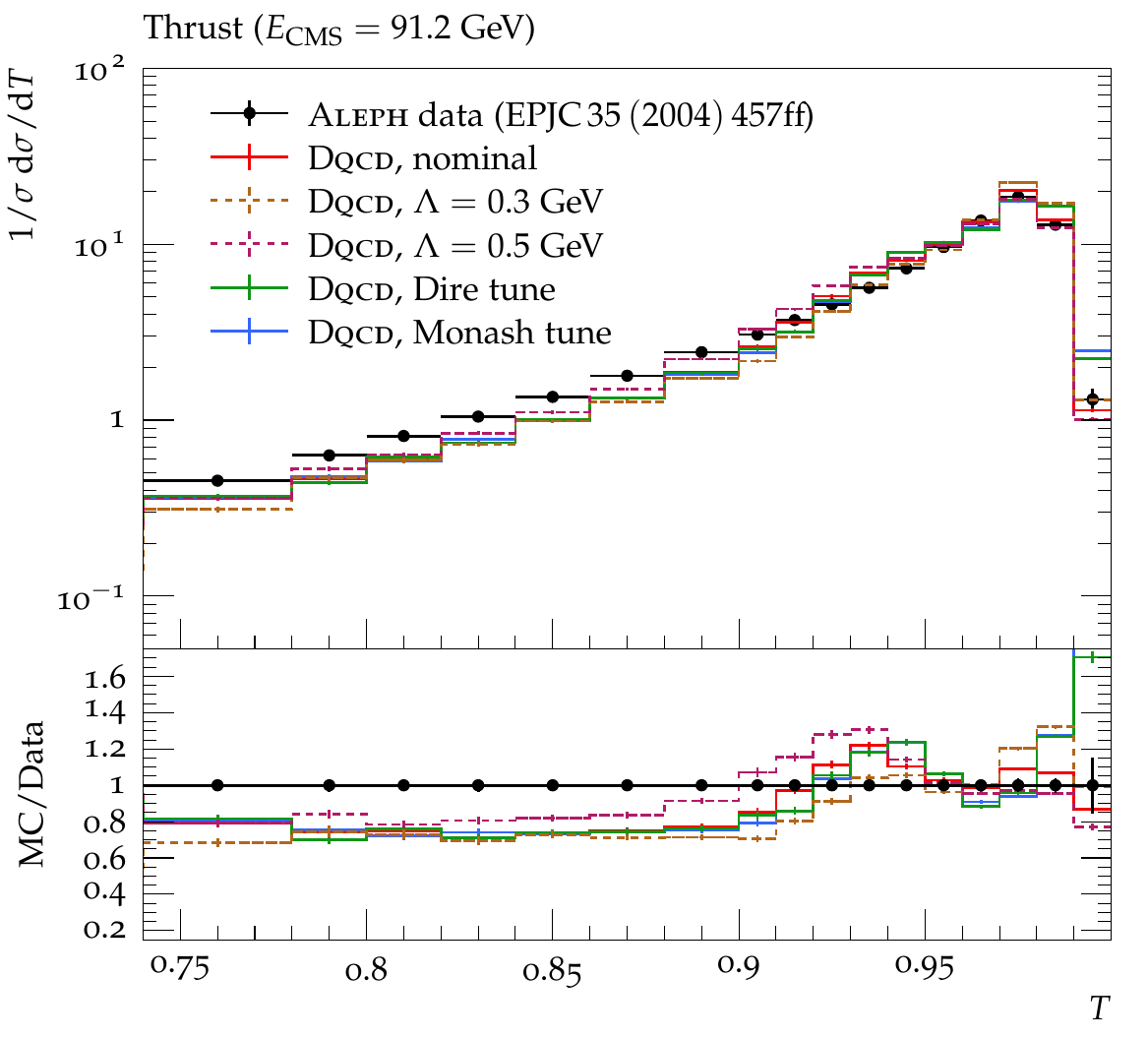}{}\\
\includegraphics[width=0.475\textwidth]{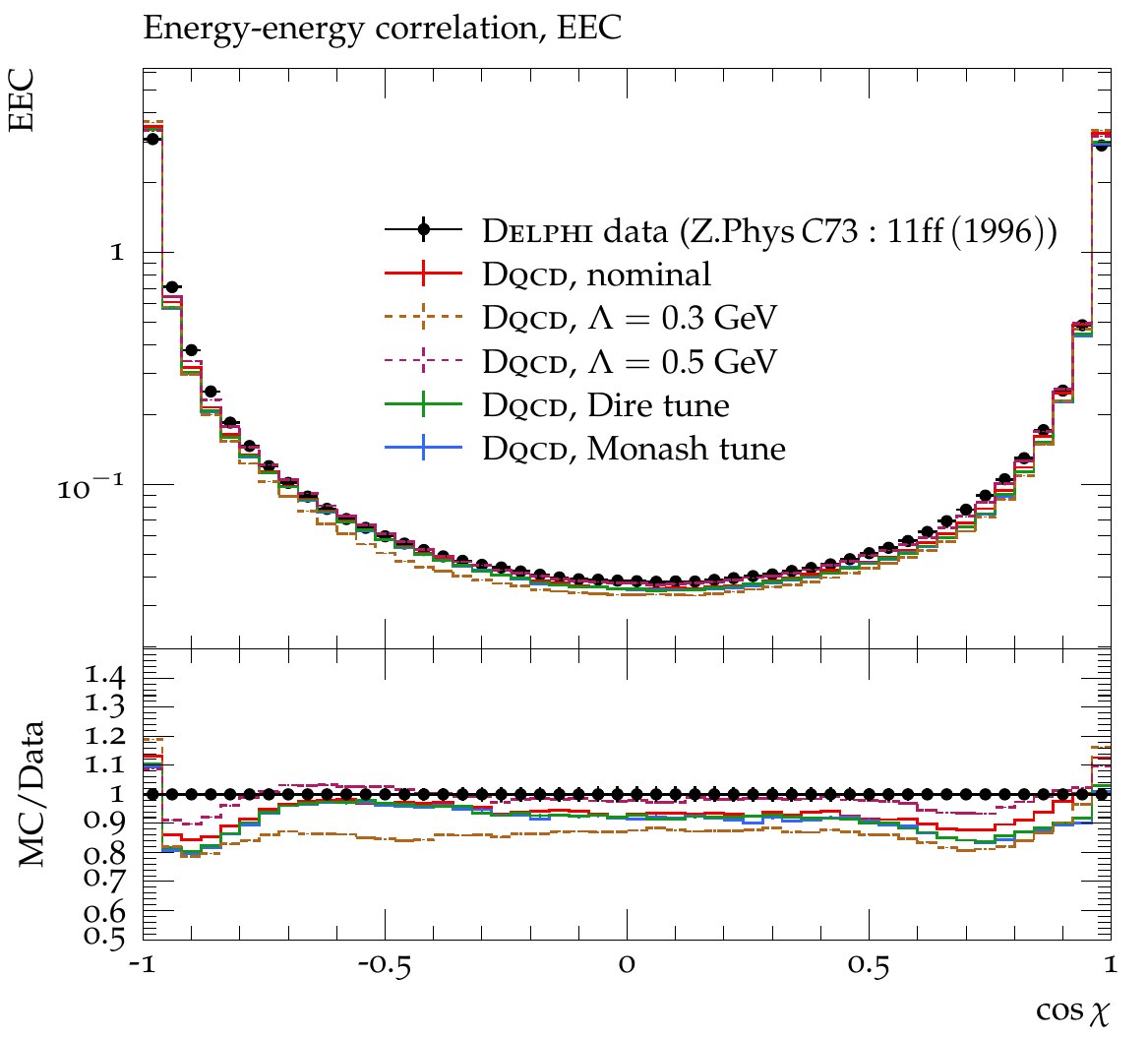}{}
\includegraphics[width=0.475\textwidth]{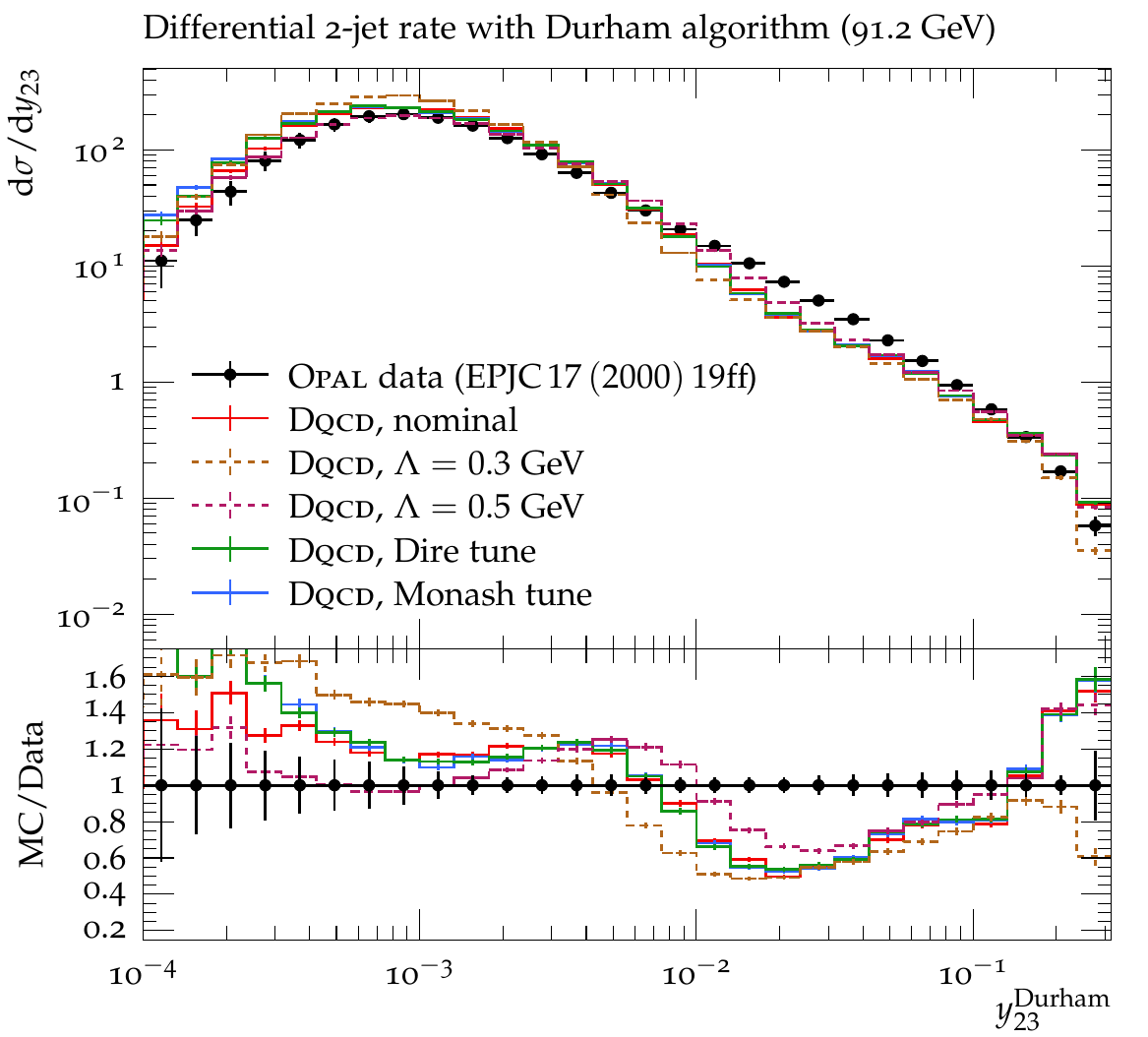}{}
\caption{Sample comparisons of the Discrete QCD model for varying values for the mass scale $\Lambda$. This leads to non-negligible uncertainties for observables that are dominated by perturbative QCD, as expected from a leading-logarithmic model.\label{fig:suppl-lep-comparison-pert}}
\end{figure}

To mitigate these errors efficiently, an in-depth study of the transpilation process needs to be performed using the IBM Q network~\cite{IBMQ}. This allows the user to control the level of optimisation of the circuit to achieve a compromise between the circuit depth and noise abatement. However, sophisticated error mitigation is not required for the example presented in this paper. Section~\ref{sec:eventGen} outlines the event generation from primitive grove structures constructed using the quantum device. Figure~\ref{fig:lep-comparison} shows that the algorithm is impressively robust to noise from the device, demonstrating the usability and practicality of NISQ devices for exciting problems in high-energy physics. 

\section{Supplementary data comparisons}\label{app:supplementary_plots}

\begin{figure}
\includegraphics[width=0.32\textwidth]{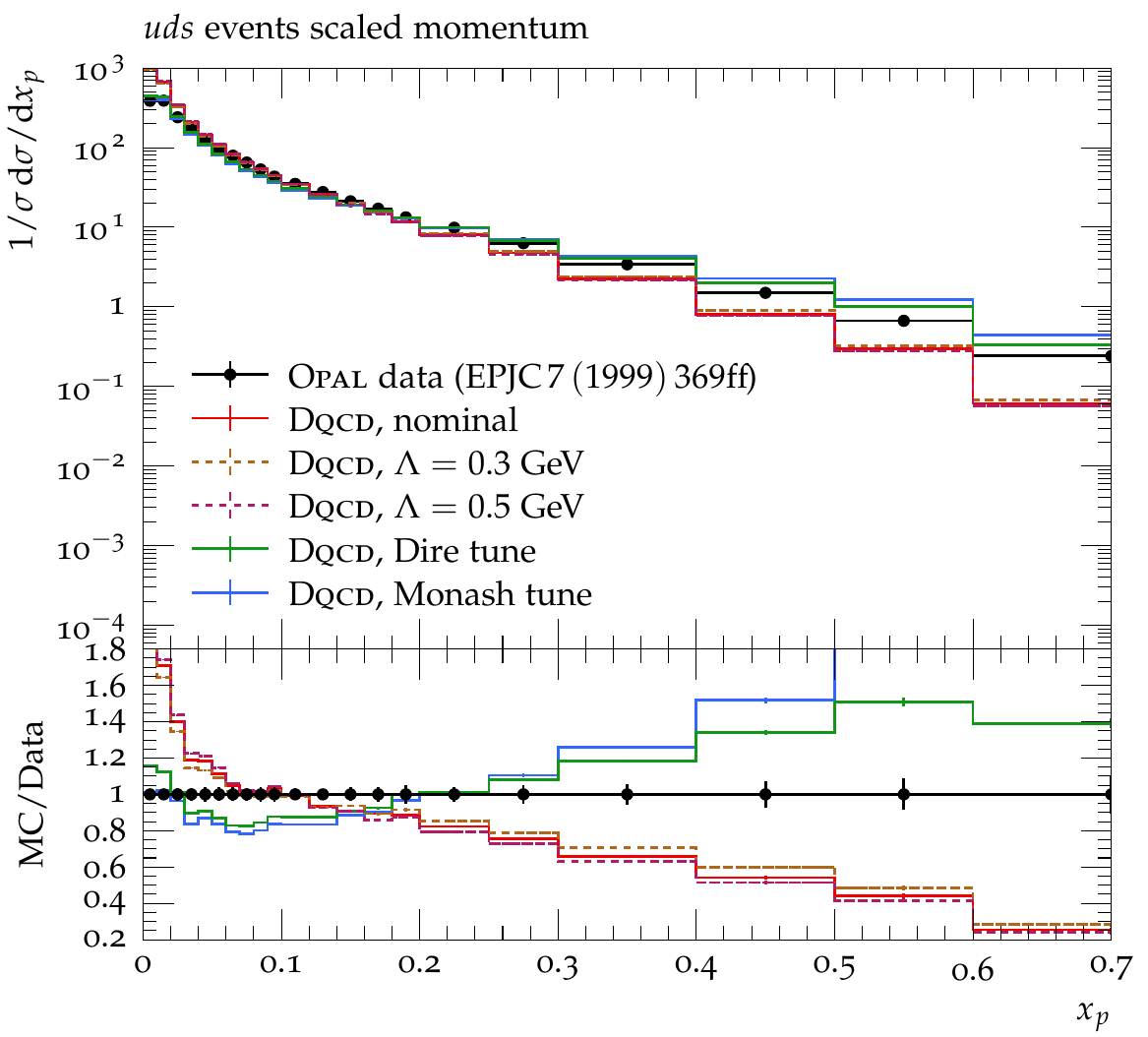}{}
\includegraphics[width=0.32\textwidth]{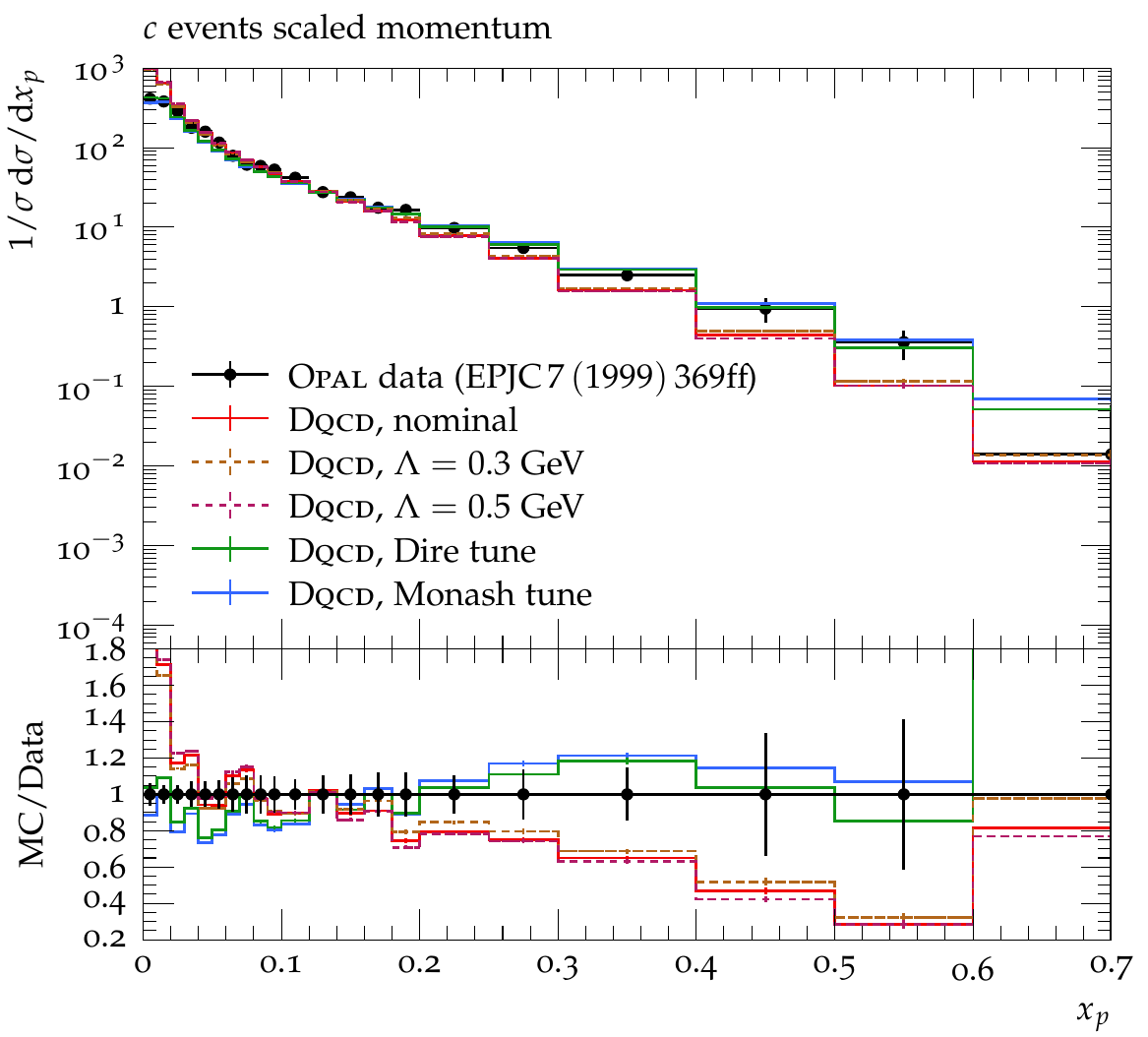}{}
\includegraphics[width=0.32\textwidth]{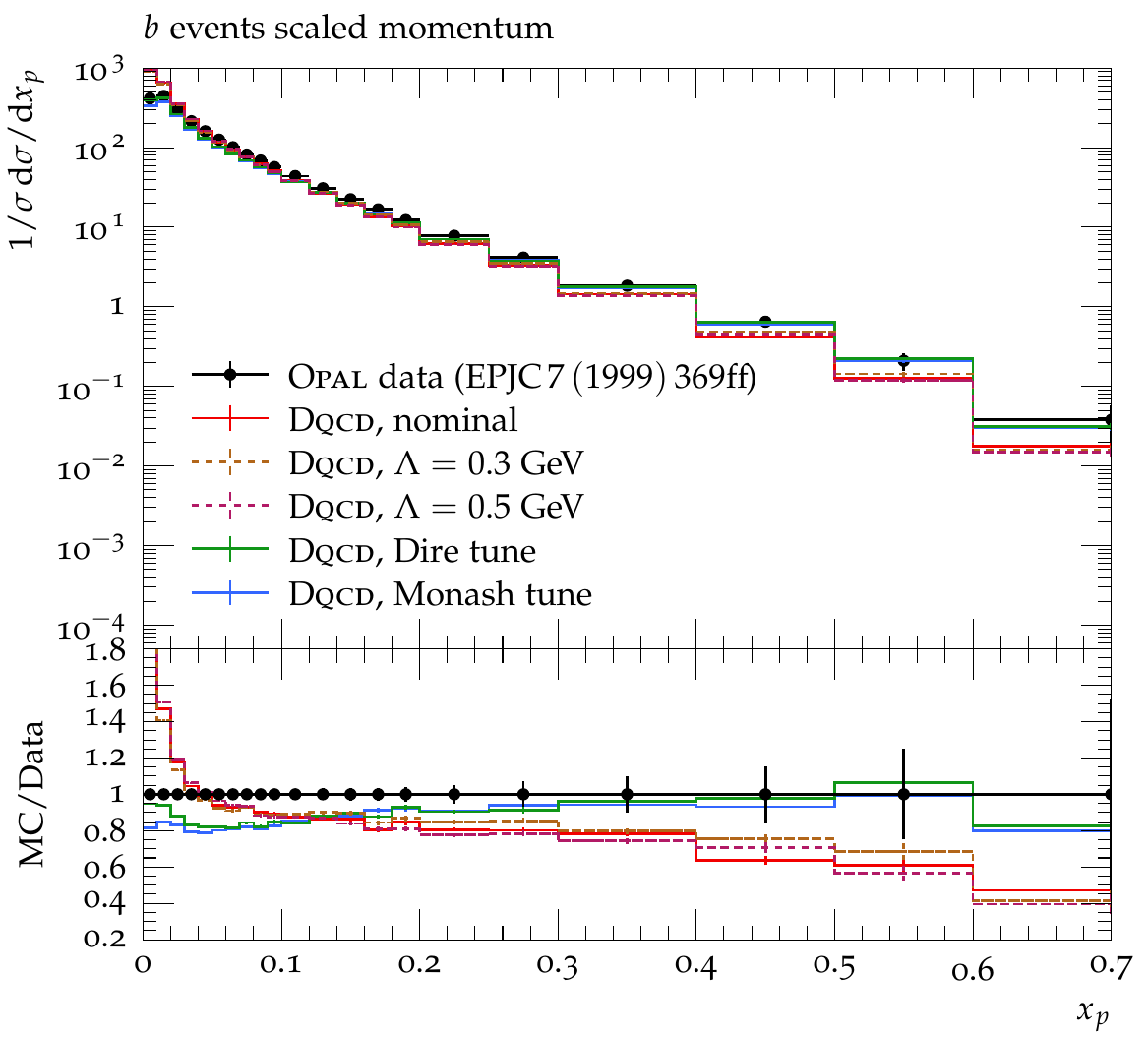}{}\\
\includegraphics[width=0.32\textwidth]{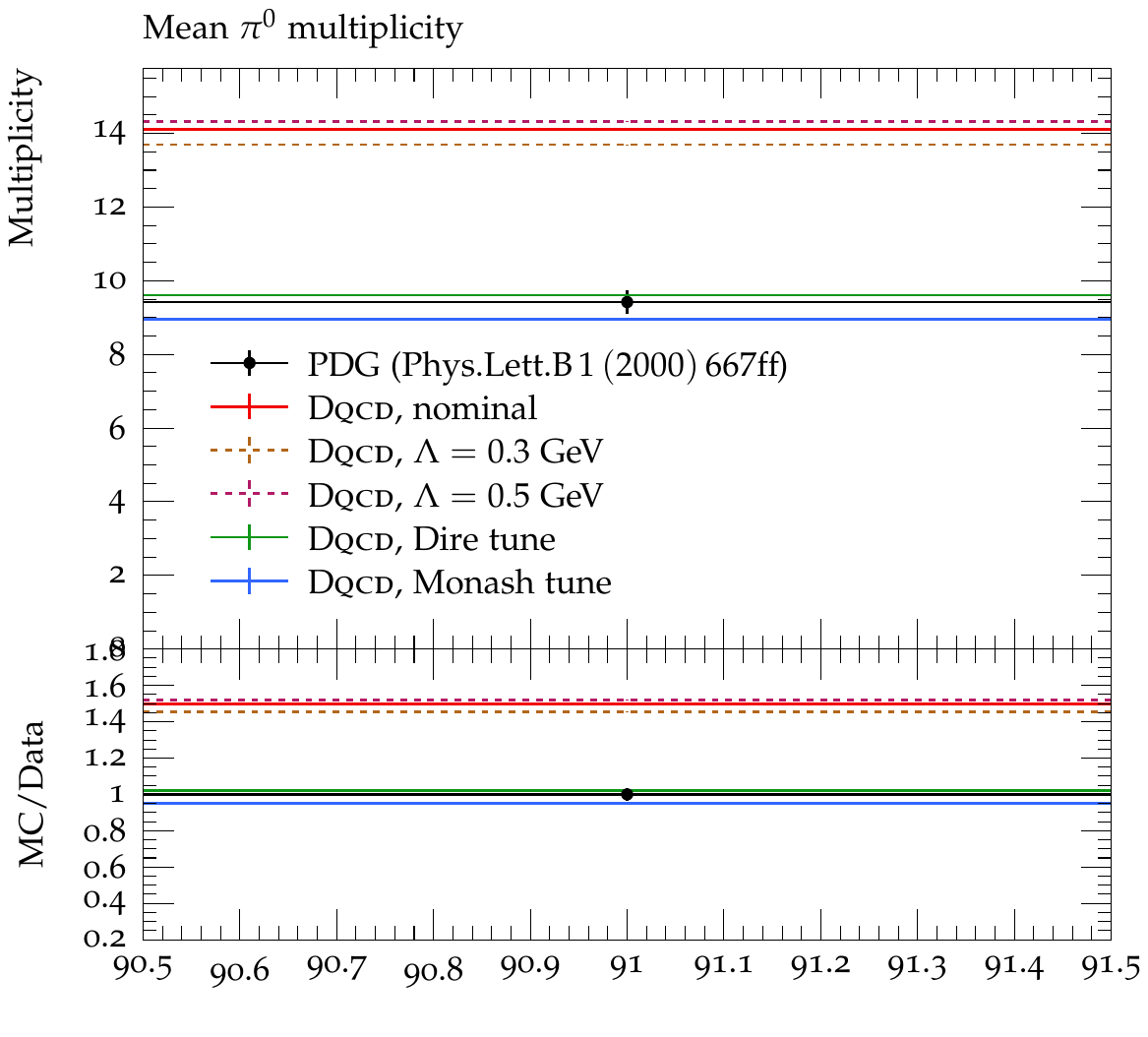}{}
\includegraphics[width=0.32\textwidth]{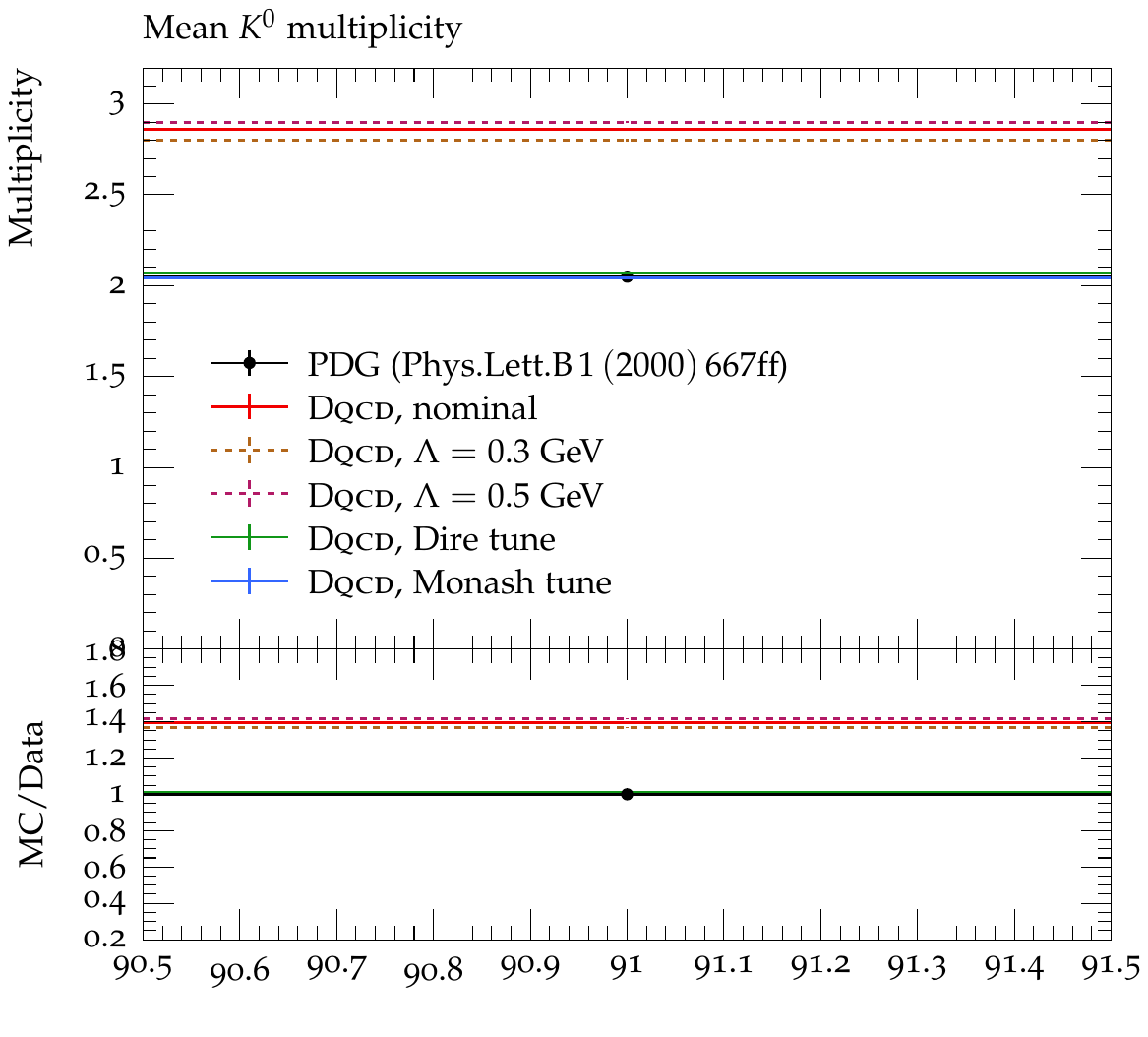}{}
\includegraphics[width=0.32\textwidth]{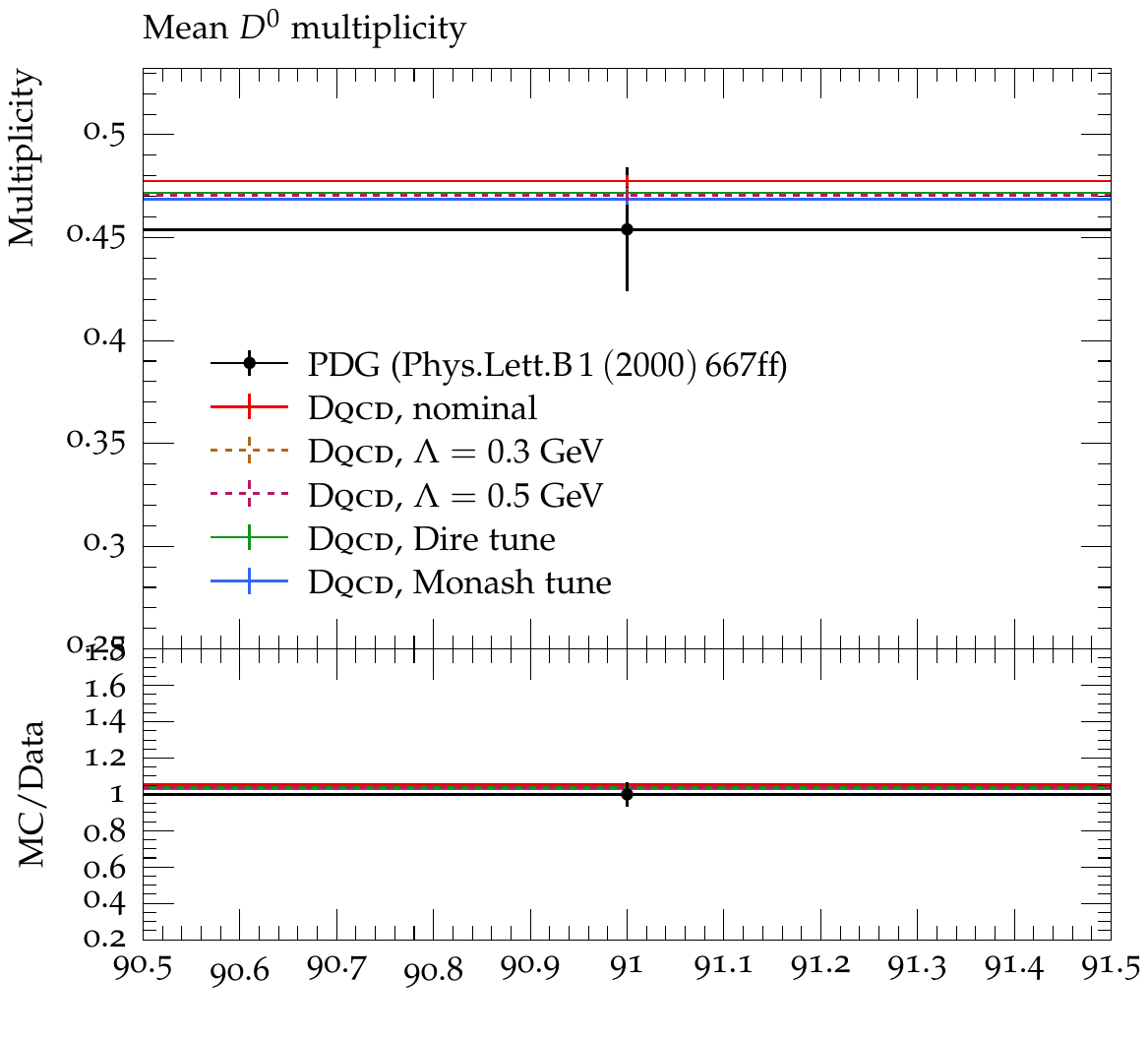}{}
\caption{Comparative study for the affect of different tunes on the data agreement. Observables dominated by non-perturbative dynamics show mild-dependance on the mass scale $\Lambda$, but are sensitive to changes in the tune.\label{fig:suppl-lep-comparison-nonpert}}
\end{figure}

The classical Discrete QCD algorithm introduced in Sec.~\ref{sec:classicalAlgo} and \ref{sec:kinematics} is one of the main developments of this article. It constitutes an compact reformulation of the orignal approach~\cite{Andersson:1995jv}. As such, additional data comparisons can be helpful to assess the predictions provided by the method.

The Discrete QCD method provides an elegant model for perturbative aspects of the showering process. Non-perturbative aspects are included by employing the Lund string model implementation of \textsc{Pythia}~\cite{Bierlich:2022pfr}. The only parameter of the perturbative model is the mass scale $\Lambda$. Figure \ref{fig:suppl-lep-comparison-pert} shows that variations of the mass scale do indeed yield non-negligible uncertainties, as should be expected from a leading-logarithmic model. Changes of the non-perturbative model (by varying the \textsc{Pythia} tune) lead to smaller variations. Examples of observables that are dominated by non-perturbative dynamics are shown in Fig.~\ref{fig:suppl-lep-comparison-nonpert}. These exhibit only very mild dependence on the value of $\Lambda$, but are very sensitive to changes in the tune. Both the default \textsc{Pythia} ``Monash" tune~\cite{Skands:2014pea} as well as the \textsc{Dire} parton shower tune~\cite{Isaacson:2018zdi} yield a better agreement with data than the nominal parameters employed for the \textsc{Dqcd} shower. This confirms a dedicated tuning of the \textsc{Dqcd} shower could improve the overall description of data. We have refrained from this, to not give an artificially positive impression of the \textsc{Dqcd} shower results. 

\bibliography{ref}{}

\begin{thebibliography}{10}

\bibitem{Fox:1979ag}
G.~C. Fox and S.~Wolfram,
\newblock Nucl. Phys. B {\bf 168}, 285 (1980).

\bibitem{Sjostrand:1985xi}
T.~Sjostrand,
\newblock Phys. Lett. B {\bf 157}, 321 (1985).

\bibitem{Gustafson:1986db}
G.~Gustafson,
\newblock Phys. Lett. B {\bf 175}, 453 (1986).

\bibitem{Lonnblad:1992tz}
L.~Lonnblad,
\newblock Comput. Phys. Commun. {\bf 71}, 15 (1992).

\bibitem{Gieseke:2003rz}
S.~Gieseke, P.~Stephens, and B.~Webber,
\newblock JHEP {\bf 12}, 045 (2003), hep-ph/0310083.

\bibitem{Sjostrand:2004ef}
T.~Sjostrand and P.~Z. Skands,
\newblock Eur. Phys. J. C {\bf 39}, 129 (2005), hep-ph/0408302.

\bibitem{Schumann:2007mg}
S.~Schumann and F.~Krauss,
\newblock JHEP {\bf 03}, 038 (2008), 0709.1027.

\bibitem{Platzer:2009jq}
S.~Platzer and S.~Gieseke,
\newblock JHEP {\bf 01}, 024 (2011), 0909.5593.

\bibitem{Nagy:2014mqa}
Z.~Nagy and D.~E. Soper,
\newblock JHEP {\bf 06}, 097 (2014), 1401.6364.

\bibitem{Hoche:2015sya}
S.~H\"oche and S.~Prestel,
\newblock Eur. Phys. J. C {\bf 75}, 461 (2015), 1506.05057.

\bibitem{Fischer:2016vfv}
N.~Fischer, S.~Prestel, M.~Ritzmann, and P.~Skands,
\newblock Eur. Phys. J. C {\bf 76}, 589 (2016), 1605.06142.

\bibitem{Hamilton:2020rcu}
K.~Hamilton, R.~Medves, G.~P. Salam, L.~Scyboz, and G.~Soyez,
\newblock JHEP {\bf 03}, 041 (2021), 2011.10054.

\bibitem{Feynman:1981tf}
R.~P. Feynman,
\newblock Int. J. Theor. Phys. {\bf 21}, 467 (1982).

\bibitem{Jordan:2014tma}
S.~P. Jordan, K.~S.~M. Lee, and J.~Preskill,
\newblock (2014), 1404.7115.

\bibitem{Garcia-Alvarez:2014uda}
L.~Garc\'\i{}a-\'Alvarez {\em et~al.},
\newblock Phys. Rev. Lett. {\bf 114}, 070502 (2015), 1404.2868.

\bibitem{Arrighi:2018PRA}
P.~Arrighi, G.~D. Molfetta, I.~M{\'a}rquez-M{\'a}rtin, and A.~P{\'e}rez,
\newblock Physical Review A {\bf 97}, 062111 (2018), 1803.01015.

\bibitem{Marque-Martin:2018PRA}
I.~M{\'a}rquez-M{\'a}rtin, P.~Arnault, G.~D. Molfetta, and A.~P{\'e}rez,
\newblock Physical Review A {\bf 98}, 032333 (2018), 1808.04488.

\bibitem{Alexandru:2019nsa}
NuQS, A.~Alexandru {\em et~al.},
\newblock Phys. Rev. D {\bf 100}, 114501 (2019), 1906.11213.

\bibitem{Jay:2019PRA}
G.~Jay, F.~Debbasch, and J.~B. Wang,
\newblock Physical Review A {\bf 99}, 032113 (2019), 1803.01304.

\bibitem{Wei:2019rqy}
A.~Y. Wei, P.~Naik, A.~W. Harrow, and J.~Thaler,
\newblock Phys. Rev. D {\bf 101}, 094015 (2020), 1908.08949.

\bibitem{Lamm:2019bik}
NuQS, H.~Lamm, S.~Lawrence, and Y.~Yamauchi,
\newblock Phys. Rev. D {\bf 100}, 034518 (2019), 1903.08807.

\bibitem{MottQuantum}
A.~Mott, J.~Job, J.-R. Vlimant, D.~Lidar, and M.~Spiropulu,
\newblock Nature {\bf 550}, 375 (2017).

\bibitem{Bauer:2019qxa}
C.~W. Bauer, W.~A. De~Jong, B.~Nachman, and D.~Provasoli,
\newblock (2019), 1904.03196.

\bibitem{Alexandru:2019ozf}
NuQS, A.~Alexandru, P.~F. Bedaque, H.~Lamm, and S.~Lawrence,
\newblock Phys. Rev. Lett. {\bf 123}, 090501 (2019), 1903.06577.

\bibitem{Blance:2020ktp}
A.~Blance and M.~Spannowsky,
\newblock JHEP {\bf 21}, 170 (2020), 2103.03897.

\bibitem{Lamm:2019uyc}
NuQS, H.~Lamm, S.~Lawrence, and Y.~Yamauchi,
\newblock Phys. Rev. Res. {\bf 2}, 013272 (2020), 1908.10439.

\bibitem{Abel:2020ebj}
S.~Abel, N.~Chancellor, and M.~Spannowsky,
\newblock (2020), 2003.07374.

\bibitem{Abel:2020qzm}
S.~Abel and M.~Spannowsky,
\newblock PRX Quantum {\bf 2}, 010349 (2021), 2006.06003.

\bibitem{Bepari:2020xqi}
K.~Bepari, S.~Malik, M.~Spannowsky, and S.~Williams,
\newblock Phys. Rev. D {\bf 103}, 076020 (2021), 2010.00046.

\bibitem{Williams:2021lvr}
S.~Williams, S.~Malik, M.~Spannowsky, and K.~Bepari,
\newblock (2021), 2109.13975.

\bibitem{DiMolfetta:2020QIP}
G.~D. Molfetta and P.~Arrighi,
\newblock Quantum Inf Process {\bf 19}, 47 (2020), 1906.04483.

\bibitem{Araz:2022haf}
J.~Y. Araz and M.~Spannowsky,
\newblock (2022), 2202.10471.

\bibitem{Matchev:2020wwx}
K.~T. Matchev, P.~Shyamsundar, and J.~Smolinsky,
\newblock (2020), 2003.02181.

\bibitem{DeJong:2020riy}
W.~A. De~Jong {\em et~al.},
\newblock Phys. Rev. D {\bf 104}, 051501 (2021), 2010.03571.

\bibitem{Ngairangbam:2021yma}
V.~S. Ngairangbam, M.~Spannowsky, and M.~Takeuchi,
\newblock Phys. Rev. D {\bf 105}, 095004 (2022), 2112.04958.

\bibitem{Bauer:2022hpo}
C.~W. Bauer {\em et~al.},
\newblock (2022), 2204.03381.

\bibitem{Agliardi:2022ghn}
G.~Agliardi, M.~Grossi, M.~Pellen, and E.~Prati,
\newblock Phys. Lett. B {\bf 832}, 137228 (2022), 2201.01547.

\bibitem{Buckley:2011ms}
A.~Buckley {\em et~al.},
\newblock Phys. Rept. {\bf 504}, 145 (2011), 1101.2599.

\bibitem{Deliyannis:2022uyh}
P.~Deliyannis {\em et~al.},
\newblock (2022), 2203.10018.

\bibitem{Field:1977fa}
R.~D. Field and R.~P. Feynman,
\newblock Nucl. Phys. B {\bf 136}, 1 (1978).

\bibitem{Collins:1984kg}
J.~C. Collins, D.~E. Soper, and G.~F. Sterman,
\newblock Nucl. Phys. B {\bf 250}, 199 (1985).

\bibitem{Collins:1989gx}
J.~C. Collins, D.~E. Soper, and G.~F. Sterman,
\newblock Adv. Ser. Direct. High Energy Phys. {\bf 5}, 1 (1989),
  hep-ph/0409313.

\bibitem{PhysRevA.48.1687}
Y.~Aharonov, L.~Davidovich, and N.~Zagury,
\newblock Phys. Rev. A {\bf 48}, 1687 (1993).

\bibitem{QWProc}
D.~Aharonov, A.~Ambainis, J.~Kempe, and U.~Vazirani,
\newblock STOC '01: Proceedings of the thirty-third annual ACM symposium on Theory of computing, pg. 50 (2001).

\bibitem{Kempe}
J.~Kempe,
\newblock Contemporary Physics {\bf 44}, 307 (2003)
  
\bibitem{Sjostrand:2006za}
T.~Sjostrand, S.~Mrenna, and P.~Z. Skands,
\newblock JHEP {\bf 05}, 026 (2006), hep-ph/0603175.

\bibitem{Bierlich:2022pfr}
C.~Bierlich {\em et~al.},
\newblock (2022), 2203.11601.

\bibitem{Azimov:1985zta}
Y.~I. Azimov, Y.~L. Dokshitzer, V.~A. Khoze, and S.~I. Troian,
\newblock Phys. Lett. B {\bf 165}, 147 (1985).

\bibitem{Gustafson:1987rq}
G.~Gustafson and U.~Pettersson,
\newblock Nucl. Phys. B {\bf 306}, 746 (1988).

\bibitem{Andersson:1995jv}
B.~Andersson, G.~Gustafson, and J.~Samuelsson,
\newblock Nucl. Phys. B {\bf 463}, 217 (1996).

\bibitem{Gehrmann-DeRidder:2011gkt}
A.~Gehrmann-De~Ridder, M.~Ritzmann, and P.~Z. Skands,
\newblock Phys. Rev. D {\bf 85}, 014013 (2012), 1108.6172.

\bibitem{McGettrick2010}
M.~McGettrick,
\newblock Quant. Inf. and Comput. , 509 (2010).

\bibitem{Shakeel2014}
A.~Shakeel, D.~Meyer, and P.~Love,
\newblock J. Math. Phys. {\bf 55}, 122204 (2014).

\bibitem{Camilleri2014}
E.~Camilleri, P.~Rohde, and J.~Twamley,
\newblock Sci Rep {\bf 4}, 4791 (2014).

\bibitem{PhysRevA.67.052317}
T.~A. Brun, H.~A. Carteret, and A.~Ambainis,
\newblock Phys. Rev. A {\bf 67}, 052317 (2003).

\bibitem{PhysRevA.87.052302}
P.~P. Rohde, G.~K. Brennen, and A.~Gilchrist,
\newblock Phys. Rev. A {\bf 87}, 052302 (2013).

\bibitem{PhysRevA.93.042323}
D.~Li, M.~Mc~Gettrick, F.~Gao, J.~Xu, and Q.-Y. Wen,
\newblock Phys. Rev. A {\bf 93}, 042323 (2016).

\bibitem{Roget2020}
M.~Roget, B.~Herzog, and G.~Di~Molfetta,
\newblock Scientific Reports {\bf 10}, 2045 (2020).

\bibitem{PhysRevLett.101.130504}
R.~D. Somma, S.~Boixo, H.~Barnum, and E.~Knill,
\newblock Phys. Rev. Lett. {\bf 101}, 130504 (2008).

\bibitem{PhysRevA.78.042336}
P.~Wocjan and A.~Abeyesinghe,
\newblock Phys. Rev. A {\bf 78}, 042336 (2008).

\bibitem{PhysRevA.67.052307}
N.~Shenvi, J.~Kempe, and K.~B. Whaley,
\newblock Phys. Rev. A {\bf 67}, 052307 (2003).

\bibitem{Szegedy}
M.~Szegedy,
\newblock Quantum speed-up of markov chain based algorithms,
\newblock in {\em 45th Annual IEEE Symposium on Foundations of Computer
  Science}, pp. 32--41, 2004.

\bibitem{Montanaro}
A.~Montanaro,
\newblock Proc. R. Soc. {\bf 471} (2015).

\bibitem{Lemieux2020efficientquantum}
J.~Lemieux, B.~Heim, D.~Poulin, K.~Svore, and M.~Troyer,
\newblock {Quantum} {\bf 4}, 287 (2020).

\bibitem{Levin2008}
D.~Levin, Y.~Peres, and E.~Wilmer,
\newblock {\em Markov Chains and Mixing Times} (American Mathematical Soc.,
  2008).

\bibitem{PhysRevA.76.042306}
P.~C. Richter,
\newblock Phys. Rev. A {\bf 76}, 042306 (2007).

\bibitem{Orsucci2018fasterquantummixing}
D.~Orsucci, H.~J. Briegel, and V.~Dunjko,
\newblock {Quantum} {\bf 2}, 105 (2018).

\bibitem{PhysRevA.104.032215}
Y.~Atia and S.~Chakraborty,
\newblock Phys. Rev. A {\bf 104}, 032215 (2021).

\bibitem{ALEPH:2003obs}
ALEPH, A.~Heister {\em et~al.},
\newblock Eur. Phys. J. C {\bf 35}, 457 (2004).

\bibitem{DELPHI:1996sen}
DELPHI, P.~Abreu {\em et~al.},
\newblock Z. Phys. C {\bf 73}, 11 (1996).

\bibitem{JADE:1999zar}
JADE, OPAL, P.~Pfeifenschneider {\em et~al.},
\newblock Eur. Phys. J. C {\bf 17}, 19 (2000), hep-ex/0001055.

\bibitem{Alwall:2006yp}
J.~Alwall {\em et~al.},
\newblock Comput. Phys. Commun. {\bf 176}, 300 (2007), hep-ph/0609017.

\bibitem{Andersen:2014efa}
J.~R. Andersen {\em et~al.},
\newblock (2014), 1405.1067.

\bibitem{Andersson:1983ia}
B.~Andersson, G.~Gustafson, G.~Ingelman, and T.~Sjostrand,
\newblock Phys. Rept. {\bf 97}, 31 (1983).

\bibitem{Buckley:2010ar}
A.~Buckley {\em et~al.},
\newblock Comput. Phys. Commun. {\bf 184}, 2803 (2013), 1003.0694.

\bibitem{Bierlich:2019rhm}
C.~Bierlich {\em et~al.},
\newblock SciPost Phys. {\bf 8}, 026 (2020), 1912.05451.

\bibitem{Andersson:1988ee}
B.~Andersson, P.~Dahlkvist, and G.~Gustafson,
\newblock Phys. Lett. B {\bf 214}, 604 (1988).

\bibitem{Klco:2018kyo}
N.~Klco {\em et~al.},
\newblock Phys. Rev. A {\bf 98}, 032331 (2018), 1803.03326.

\bibitem{Gustafson:1992uh}
G.~Gustafson,
\newblock Nucl. Phys. B {\bf 392}, 251 (1993).


\bibitem{Frixione:2002ik}
S.~Frixione and B.~R. Webber,
\newblock JHEP {\bf 06}, 029 (2002), hep-ph/0204244.

\bibitem{Nason:2004rx}
P.~Nason,
\newblock JHEP {\bf 11}, 040 (2004), hep-ph/0409146.

\bibitem{Frixione:2007vw}
S.~Frixione, P.~Nason, and C.~Oleari,
\newblock JHEP {\bf 11}, 070 (2007), 0709.2092.

\bibitem{Catani:2001cc}
S.~Catani, F.~Krauss, R.~Kuhn, and B.~R. Webber,
\newblock JHEP {\bf 11}, 063 (2001), hep-ph/0109231.

\bibitem{Lonnblad:2001iq}
L.~L{\"o}nnblad,
\newblock JHEP {\bf 05}, 046 (2002), hep-ph/0112284.

\bibitem{Mangano:2001xp}
M.~L. Mangano, M.~Moretti, and R.~Pittau,
\newblock Nucl. Phys. {\bf B632}, 343 (2002), hep-ph/0108069.

\bibitem{Mrenna:2003if}
S.~Mrenna and P.~Richardson,
\newblock JHEP {\bf 05}, 040 (2004), hep-ph/0312274.

\bibitem{Alwall:2007fs}
J.~Alwall {\em et~al.},
\newblock Eur. Phys. J. C {\bf 53}, 473 (2008), 0706.2569.

\bibitem{RevModPhys.82.1155}
A.~A. Clerk, M.~H. Devoret, S.~M. Girvin, F.~Marquardt, and R.~J. Schoelkopf,
\newblock Rev. Mod. Phys. {\bf 82}, 1155 (2010).

\bibitem{FrontEng}
N.~A. of~Engineering~2019,
\newblock {\em Frontiers of Engineering: Reports on Leading-Edge Engineering
  from the 2018 Symposium} (The National Academies Press, 2018).

\bibitem{lidar2013quantum}
D.~Lidar and T.~Brun,
\newblock {\em Quantum Error Correction} (Cambridge University Press, 2013).

\bibitem{Devitt_2013}
S.~J. Devitt, W.~J. Munro, and K.~Nemoto,
\newblock Reports on Progress in Physics {\bf 76}, 076001 (2013).

\bibitem{QiskitT1}
Calibrating qubits using qiskit pulse, https://learn.qiskit.org/course/quantum-hardware-pulses/calibrating-qubits-using-qiskit-pulse

\bibitem{IBMQ}
IBM Quantum, 2021.

\bibitem{Skands:2014pea}
P.~Skands, S.~Carrazza and J.~Rojo,
\newblock Eur. Phys. J. C \textbf{74} (2014) no.8, 3024

\bibitem{Isaacson:2018zdi}
J.~Isaacson and S.~Prestel,
\newblock Phys. Rev. D \textbf{99} (2019) no.1, 014021



\end{thebibliography}
\bibliographystyle{h-physrev}

\end{document}